\documentstyle[12pt,aaspp4]{article}

\def\O{\Omega}
\def\te{T_{eff}}
\def\cm{{\rm\ cm}} 
\def\mic{\mu {\rm m}}
\def\ang{\AA\ }
\def\xec{\times 10^{11} {\rm\ cm}}

\def\kms{{~{\rm km~s}^{-1}}} 
\def\pers{~{\rm s}^{-1}} 
\def\xfs{\times 10^{-5} \pers}
\def\xss{\times 10^{-6} \pers}
\def\ergs{{\rm ~ergs~s^{-1}}}

\def\lfl{\lambda F_\lambda}
\def\msun{{\,\rm M_{\odot}}}
\def\rsun{{\,\rm R_{\odot}}}
\def\lsun{{\,\rm L_{\odot}}}
\def\lacc{\,L_{acc}}
\def\msyr{{\msun}~{\rm yr}^{-1}}
\def\O{\Omega}

\def\ok{\Omega_K}
\def\oms{\Omega_*} 

\def\omks{\Omega_K(R_{_*})}
\def\omk2i{\Omega_K^2(R_{in})}
\def\omk2o{\Omega_K^2(R_{out})}
\def\omk2s{\Omega_K^2(R_{_*})}
\def\md{\dot M}
\def\ms{M_{_*}}
\def\rs{R_{_*}}

\def\ro{R_{out}}
\def\hs{H_{_*}}
\def\vs{v_{R,*}}

\def\et{{\it et al.}\ }

\def\sles{\lower2pt\hbox{$\buildrel {\scriptstyle <}
   \over {\scriptstyle\sim}$}}
\def\sgreat{\lower2pt\hbox{$\buildrel {\scriptstyle >}
   \over {\scriptstyle\sim}$}}

\begin{document}

\title{SPECTRA AND LINE PROFILES OF FU ORIONIS OBJECTS: COMPARISONS
BETWEEN BOUNDARY LAYER MODELS AND OBSERVATIONS}
\author{Robert Popham, Scott Kenyon, Lee Hartmann, and Ramesh Narayan}
\affil{Harvard-Smithsonian Center for Astrophysics}
\authoraddr{MS 51, 60 Garden St., Cambridge, MA 02138}

\begin{abstract}

We present solutions for the accretion disks and boundary layers in
pre-main-sequence stars undergoing FU Orionis outbursts. These
solutions differ from earlier disk solutions in that they include a
self-consistent treatment of the boundary layer region. In a previous
paper (Popham 1996), we showed that these stars should stop accreting
angular momentum once they spin up to modest rotation rates. Here we
show that for reasonable values of $\alpha$, these low angular
momentum accretion rate solutions fit the spectra and line profiles
observed in FU Orionis objects better than solutions with high rates
of angular momentum accretion. We find solutions which fit the
observations of FU Orionis and V1057 Cygni. These solutions have mass
accretion rates of 2 and $1 \times 10^{-4} \msyr$, stellar masses of
0.7 and $0.5 \msun$, and stellar radii of 5.75 and $5.03 \rsun$,
respectively. They also have modest stellar rotation rates $8-9 \xss$,
comparable to the observed rotation rates of T Tauri stars, and
angular momentum accretion rates of zero. This supports our earlier
suggestion that FU Orionis outbursts may regulate the rotation rates
of T Tauri stars.

\end{abstract}


\section{Introduction}

The FU Orionis objects are a class of accreting pre-main-sequence
stars which are undergoing dramatic outbursts in their brightness (see
Hartmann, Kenyon, \& Hartigan 1993; Kenyon 1995; Hartmann \& Kenyon
1996 for reviews). These outbursts are believed to result from a
sudden, large increase in the mass accretion rate in an accretion disk
around a T Tauri star. Line profiles in FU Orionis systems are
generally double-peaked, as expected from a rotating disk (Hartmann \&
Kenyon 1985, 1987). The observed variation in linewidths as a function
of temperature also suggests a disk origin; line profiles in the
optical are broader than those in the infrared. This is the trend
expected for lines produced in a Keplerian disk, where the rotational
velocity and temperature both decrease with radius, so that smaller
rotational linewidths are produced in cooler regions (Hartmann \&
Kenyon 1987). The spectral energy distributions of these systems are
also quite broad, which is consistent with a disk with a range of
effective temperatures at different radii.

Accretion disk models have been successful in reproducing observations
of individual systems. Kenyon, Hartmann, \& Hewett (1988, hereafter
KHH) constructed steady thin disk models for FU Orionis and V1057
Cygni.  They found that simple thin disk models were able to reproduce
both line profiles and broad-band spectra of these systems. These
models had mass accretion rates of about $10^{-4} \msyr$, stellar
masses of $0.3 - 1.0 \msun$, and stellar radii of $4-6 \rsun$. They
had Keplerian rotational velocities, and maximum disk temperatures of
7200 K for FU Orionis and 6590 K for V1057 Cygni.

One major simplifying assumption made by the KHH models is that the
accretion disks in FU Orionis systems can be approximated by a
standard Keplerian thin disk model. This assumption breaks down in the
inner disk, where the transition from the disk to the accreting star
occurs in the boundary layer region. Early estimates for the
temperature of the boundary layer region suggested that it would be
much hotter than permitted by observations (Hartmann \& Kenyon 1985;
Kenyon \et 1989). These estimates were made by assuming that the
boundary layer luminosity is half of the total accretion luminosity,
and that it is radiated from a small region of the inner disk, with a
radial extent of a few percent of the stellar radius. These large
luminosities radiated from such a small area would produce boundary
layer temperatures approaching 30,000 K, which would mean that the
boundary layer would emit large ultraviolet fluxes. IUE observations
of FU Orionis objects by Kenyon \et (1989) demonstrated that these
large fluxes were not present. Kenyon \et (1989) suggested three
possible reasons why the boundary layer might not be as hot as
predicted by these simple estimates. First, some of the boundary layer
energy might be going into expanding the central star. Second, the
central star might be rotating rapidly due to accretion spinup. Third,
the disk might not be thin, and so the radial extent of the boundary
layer might be much larger than a few percent of the stellar radius.

Recently, we have begun to calculate self-consistent solutions for the
structure of boundary layers in FU Orionis systems. These solutions
suggest that all of the mechanisms mentioned by Kenyon \et (1989) act
to decrease the effective temperature of the boundary layer below the
temperatures suggested by simple estimates. Popham \et (1993,
hereafter PNHK) obtained boundary layer solutions for
pre-main-sequence stellar parameters and mass accretion rates $\md$
ranging from $10^{-7} - 10^{-4} \msyr$. They showed that at the higher
accretion rates, which correspond to FU Orionis systems, the radial
extent of the boundary layer region becomes quite large, and the
boundary layer temperature drops much closer to that of the inner
disk. Thus as $\md$ increases it becomes progressively more difficult
to distinguish the contribution of the boundary layer region to the
overall spectrum. PNHK also showed that in high-$\md$ systems, the
accreting material is quite hot when it reaches the stellar radius,
with a temperature which is a substantial fraction of the virial
temperature. This means that a substantial fraction of the accretion
luminosity is advected into the star, decreasing the luminosity
radiated by the boundary layer. PNHK calculated two models with $\md =
10^{-4} \msyr$, one with $\alpha = 10^{-1}$ and the other with $\alpha
= 10^{-3}$. The $\alpha = 10^{-1}$ solution reached a peak $\te \simeq
17,000$ K, with $H/R \sim 0.2$, while the $\alpha = 10^{-3}$ solution
had a lower peak $\te \simeq 9500$ K and $H/R \sim 0.4$. Godon (1996)
confirmed these results with a time-dependent model, finding a peak
$\te$ of 14,000 K and $H/R \simeq 0.40$ for $\md = 10^{-4} \msyr$ and
$\alpha = 0.3$.

Recently, Popham (1996, hereafter Paper I) presented boundary layer
solutions for FU Orionis parameters, and examined the effect of FU
Orionis outbursts on the spin evolution of the underlying T Tauri
stars. Event statistics suggest that the outbursts may dominate the
mass accretion onto T Tauri stars (Hartmann \& Kenyon 1996); if so,
they should also dominate the angular momentum accretion. Disk
accretion generally adds angular momentum to the accreting star and
spins it up; however, when the accreting star nears breakup, the
angular momentum accretion rate drops rapidly and becomes negative, so
that the accreting star stops spinning up (Popham \& Narayan 1991;
Paczy\'nski 1991). Paper I showed that for the disk parameters which
characterize FU Orionis systems, the accreting star stops spinning up
once it reaches 20-40\% of breakup speed (i.e., the Keplerian rotation
rate at the stellar surface). The star can then continue to accrete
mass while maintaining an equilibrium rotation rate. Paper I suggested
that these low angular momentum accretion rate solutions, obtained
during outbursts, play an important role in maintaining the observed
slow rotation rates of T Tauri stars.

The boundary layer region should make an important contribution to the
observed spectral energy distribution and spectral line profiles of FU
Orionis systems. In the boundary layer, the rotational velocity of the
accreting material, which determines the linewidths, varies rapidly
with radius. It drops from its disk value, generally close to
Keplerian, to the surface rotational velocity of the star, which is
usually much slower. As the material loses its rotational kinetic
energy, up to half of the accretion luminosity is released in the
boundary layer region.  Since disk models which include the boundary
layer will have quite different temperature and rotational velocity
profiles from Keplerian thin disk models, they will produce somewhat
different spectra and line profiles. Also, as shown in Paper I, low
angular momentum accretion rate solutions have different temperature
and rotational velocity profiles from those with high angular momentum
accretion rates. This raises several questions: 1) whether solutions
which include the boundary layer region will be able to match the
observations of FU Orionis systems; 2) which types of boundary layer
solutions will match the observations, and 3) how the inclusion of the
boundary layer will change the model parameters inferred from
Keplerian disk solutions.

In this paper, we attempt to answer these questions by calculating
boundary layer solutions for FU Orionis systems, and comparing the
spectra and line profiles produced by these solutions to the
observations. In \S 2, we briefly describe our boundary layer model
and the procedures we use for calculating spectra and line profiles
from our solutions. In \S 3, we discuss the general characteristics of
the spectra and profiles produced by our boundary layer solutions. We
show how these characteristics depend on solution parameters such as
the mass accretion rate, the stellar mass, radius, and rotation rate,
the angular momentum accretion rate, and the viscosity parameter
$\alpha$. We compare our solutions to observations of the spectra and
line profiles of individual FU Orionis systems in \S 4, and discuss
our results and their implications in \S 5. \S 6 gives a summary.

\section{Boundary Layer and Disk Model}

The model we use to calculate the structure of the boundary layer and
accretion disk is identical to the one used in Paper I and very
similar to the model used in several earlier papers (Narayan \& Popham
1993; PNHK; Popham \& Narayan 1995), where it is described in
detail. The model uses the steady-state, axisymmetric slim disk
equations originally developed by Paczy\`nski and collaborators for
modeling disks around black holes (Paczy\`nski \& Bisnovatyi-Kogan
1981; Muchotrzeb \& Paczy\`nski 1982; Abramowicz \et 1988) to describe
the radial structure of the disk.  The slim disk equations dispense
with some of the assumptions of the standard thin disk model (Shakura
\& Sunyaev 1973), and include terms which are assumed to be small in
the thin disk formulation. These include the radial pressure gradient
and acceleration in the radial momentum equation, and radial
energy transport by radiation and advection. Also, while the thin disk
equations assume that the angular momentum accretion rate takes on a
standard value, the slim disk equations allow for different angular
momentum accretion rates. The additional terms in the slim disk
equations involve radial derivatives which make them differential
equations requiring a numerical solution. We use approximate relations
to describe the vertical structure of the disk.  We solve the
equations using a relaxation method, on a 1001-point radial grid which
extends from $\rs$ to $100 \rs$, with grid points concentrated in the
inner boundary layer region.

\subsection{Spectra}

Our solutions yield the effective temperature of the disk surface as a
function of radius. From this, we construct spectra by simply adding
up the contributions from each disk annulus. We use two different
methods for calculating the spectrum of each annulus. The first is
simply to assume that the disk surface emits as a blackbody at the
local effective temperature. This has the advantages of being simple
and independent of any assumptions about the vertical structure of the
disk.  The second method is to use a grid of stellar atmospheres,
using the methods described by Hartmann \& Kenyon (1985).  This should
do a better job of reproducing features such as the Balmer jump;
however, it is still only approximate, since the disk atmosphere
probably differs somewhat from a stellar atmosphere.

We assume that the only effect of inclination is to change the
projected area of the disk as seen by an observer.  Thus, the observed
flux from the disk is $F_\lambda = L_\lambda \cos i / 2 \pi d^2$,
where $L_\lambda$ is the total disk luminosity, $i$ is the inclination
angle, and $d$ is the distance. At the average value of $\langle \cos
i \rangle = 1/2$, we have the standard expression $F_\lambda =
L_\lambda / 4 \pi d^2$.

\subsection{Line Profiles}

In order to calculate spectral line profiles from our models, we begin
by calculating the line profile produced by a single annulus of the
disk. We take the profile function for a rotating disk,
$\Phi(\Delta v) = [1 - (\Delta v / v_{los})^2]^{-1/2}$, where $\Delta
v$ is the deviation in velocity from line center, and $v_{los} =
v_{rot} \sin i$ is the projected rotational velocity at inclination
angle $i$. We convolve this profile function with a Gaussian
broadening function $\exp[-(\Delta v / v_b)^2]$, where $v_b$ is the
instrumental broadening due to the resolution of the spectrograph.

Next, we sum up the line profile contributions from all disk
annuli. We weight the annuli according to their contribution to the
blackbody flux at the wavelength of the spectral line. This is
equivalent to assuming that the equivalent width of the line stays
constant as the effective temperature varies. For the lines we
consider, metal lines at 6170 \ang and CO lines at 2.2 $\mic$, this is
a reasonable assumption, except at high temperatures, where the
species in question disappear. For this reason, we set the weight to
zero when the effective temperature exceeds a cutoff value, which we
take to be 8000 K for the 6170 \ang lines and 5000 K for the 2.2
$\mic$ lines.

\section{Spectra and Line Profiles from Boundary Layer Solutions and
Their Dependence on Solution Parameters}

Our disk and boundary layer solutions have a number of free parameters
which can affect their spectra and line profiles. These include the
usual parameters used to characterize accreting systems: the mass
accretion rate $\md$, and the stellar mass and radius $\ms$ and $\rs$.
As discussed in Paper I, even when these parameters are specified,
there are still a range of solutions available, which are
characterized by different values of the stellar rotation rate $\oms$
and the disk height at the stellar radius $\hs$, which then specify
the angular momentum accretion rate $\dot J$. Finally, one can specify
the parameter $\alpha$, which characterizes the magnitude of the
viscosity.

With such a wide range of solution parameters available, we begin by
examining the effects of each parameter on our model spectra and line
profiles. We then compare these to the general observed
characteristics of FU Orionis systems in order to narrow down the
parameter space in which solutions in reasonable agreement with the
data will be found. Later in \S 4, we will attempt to find solutions
which match the observations of individual systems (FU Orionis and
V1057 Cygni).

The general observed characteristics of FU Orionis systems, which we
will use to try to limit the parameter space for our solutions, are as
follows: the line profiles are generally double-peaked, both in the
optical and the infrared (Hartmann \& Kenyon 1985, 1987). The line
widths vary widely from system to system; some are below the
instrumental resolution of $10-15 \kms$, while others have FWHM values
in excess of $100 \kms$. Such a range is not surprising, since the
linewidth depends on the projected rotational velocity $v_{rot} \sin
i$. The linewidths in the optical are broader than those in the
infrared; linewidths measured at 6170 \ang are $\sim 40\%$ larger than
those measured at 2.2 $\mu m$.

The dereddened broad-band spectra of FU Orionis systems generally have
their peak $\lfl$ values in the red part of the spectrum (KHH). The
spectra fall off quite rapidly at blue and ultraviolet wavelengths,
with $\lfl$ in the U band a factor of three or more smaller than the
peak value. The best estimates for the distances of these systems give
total luminosities of a few hundred $\lsun$ (Bell \& Lin 1994;
Hartmann \& Kenyon 1996).

\subsection{Standard Accretion Parameters: $\md$, $\ms$, and $\rs$}

The three most fundamental parameters describing an accretion disk are
the mass accretion rate $\md$ and the mass $\ms$ and radius $\rs$ of
the accreting star. In the thin disk formulation, these parameters can
be combined to derive three standard quantities which characterize the
spectra and linewidths produced by the disk: the accretion luminosity
$L_{acc} = G \md \ms / \rs$, the maximum disk temperature $T_{max} =
0.287 (G \md \ms / \sigma \rs^3)^{1/4}$, and the Keplerian rotational
velocity at $\rs$, $v_K(\rs) = (G \ms / \rs)^{1/2}$. These quantities
provide a good general guide to the variations in the spectra and line
profiles that result from variations in $\md$, $\ms$, and $\rs$. The
accretion luminosity characterizes the brightness (the vertical
position) of the spectrum, the maximum temperature controls the peak
wavelength (the horizontal position) of the spectrum, and the
Keplerian rotational velocity at $\rs$ characterizes the width of a
line profile.

For a standard thin accretion disk, the quantities $L_{acc}$,
$T_{max}$, and $v_K(\rs)$ represent the exact values of the luminosity
and the maximum effective temperature and rotational velocity of the
disk. In our models, the situation is more complex, for two
reasons. First, we include the boundary layer region in our
solutions. The effects of variations in $\md$, $\ms$ and $\rs$ upon
the boundary layer are more complex than their effects upon the disk,
and the boundary layer is also affected by other parameters such as
the rotation rate of the accreting star $\oms$ and the angular
momentum accretion rate. Second, our slim disk equations include
important terms, described in \S 2, which are not included in the
standard thin disk model. Thus, for our solutions, $L_{acc}$,
$T_{max}$, and $v_K(\rs)$ are only characteristic values which provide
a good approximate description of the outer parts of the disk, but not
of the inner portions close to the star. The radiated luminosity of
our solutions can differ substantially from $L_{acc}$, since energy
can be advected into the star, and variations in $\oms$ change the
boundary layer luminosity.  The boundary layer region is usually
hotter than $T_{max}$, and the rotational velocity of the disk is
often well below Keplerian there.

As an example of the changes in the boundary layer structure produced
by variations in $\md$ and $\rs$, we show in Figure 1 three solutions
with the parameters used in Paper I. The first solution has $\md =
10^{-4.3} \msyr$, $\rs = 2.25 \xec$, the second has $\md = 10^{-4.15}
\msyr$, $\rs = 3 \xec$, and the third $\md = 10^{-4.0} \msyr$, $\rs =
4 \xec$.  Note that these choices of parameters keep the accretion
luminosity $\lacc$ approximately constant at $\simeq 10^{36} \ergs
\simeq 250 \lsun$, since the ratio $\md/\rs$ stays nearly
constant. The maximum disk temperature $T_{max}$ and the Keplerian
rotational velocity at the stellar radius $v_K(\rs)$ both decrease as
$\md$ and $\rs$ increase. All three solutions have $\ms = 0.5 \msun$,
and $\alpha = 10^{-2}$, and all three use the boundary condition $\vs
= -1000 \cm \pers$, which will be discussed below. These are high
angular momentum accretion rate solutions. Throughout this work, we
express the angular momentum accretion rate $\dot J$ as $j \equiv \dot
J / \md \omks \rs^2$, so that $j = 1$ for a standard thin disk
solution. The solutions in Fig. 1 have $j=1.0$, 0.9, and 0.8,
respectively.

Figure 1a shows the angular velocity $\O$ for these solutions. All
three have a peak in $\O$; inside the peak, $\O$ drops to $\oms \ll
\omks$, while outside the peak, $\O$ decreases at a roughly Keplerian
rate. The main feature we wish to emphasize is that the location of
the peak changes from about $1.8 \rs$ to $2.6 \rs$ as $\md$ increases;
thus the dynamical boundary layer width doubles from $0.8 \rs$ to $1.6
\rs$ as $\md$ doubles. Another important feature is that as $\md$
increases, $\O$ drops farther below $\ok$, indicated by a dashed
line. At $\md = 10^{-4.0} \msyr$, the peak value of $\O$ is only half
of $\ok$ at that radius. This is caused by the increase in pressure
support at large values of $\md$. Figure 1b shows the effective
temperatures of these solutions. All three reach peak values just
outside of $\rs$, but the peak is much more pronounced at lower
$\md$. This is not surprising, since the dissipation of energy is much
more concentrated in the smaller boundary layer. As $\md$ increases,
the peak in $\te$ due to the boundary layer dissipation becomes
broader until it becomes difficult to distinguish it from the $\te$
profile of the disk. This broadening and cooling of the boundary layer
with increasing $\md$ was pointed out in our earlier paper (PNHK) for
$\md$ ranging from $10^{-7}$ to $10^{-4} \msyr$. Fig. 1 shows that at
the upper end of this range, a factor of two increase in $\md$ can
make a qualitative difference in the appearance of the boundary layer
region.

The 6170 \ang line profiles for these solutions are shown in Figure
1c. These profiles reflect the variations in the rotational velocities
in the inner portion of the disk. All three profiles are centrally
peaked, due to the presence of the hot, slowly rotating material in
the boundary layer region. At the lowest $\md$, the rapidly rotating
inner disk produces a double-peaked component, which is filled in by
the contribution from the more slowly rotating boundary layer, leaving
broad shoulders on the combined line profile. As the boundary layer
extends to larger radii at higher values of $\md$, the double-peaked
component becomes even more difficult to distinguish; it only appears
as wings on the centrally-peaked boundary layer profile. The width of
the profiles decreases dramatically as $\md$ and $\rs$ increase, since
the peak rotational velocity is much smaller. None of the profiles
resemble the double-peaked profiles observed in FU Orionis systems;
clearly the inclusion of the boundary layer makes a crucial difference
in the overall line profile by reducing the rotational velocities over
a wide region of the inner disk, where the disk is brightest.

Figure 1d shows blackbody spectra of these solutions.  They have
nearly the same luminosity, but the spectrum moves to longer
wavelengths as $\md$ increases: the peak $\lfl$ moves from $\sim 0.5
\mic$ to $\sim 0.75 \mic$. This change in peak wavelength is larger
than would be expected from the change in the $T_{max}$ of the disk,
due to the decrease in the boundary layer effective temperature.
Separate boundary layer and disk components are not distinguishable in
the spectrum, as they are at lower values of $\md$ (PNHK).

\subsection{Stellar Rotation Rate, Disk Height, and Angular Momentum
Accretion Rate: $\oms$, $\hs$, and $j$}

In Paper I, we showed that a wide range of boundary layer solutions
can be found for a single set of values of $\md$, $\ms$, and $\rs$,
These solutions have different values of the stellar rotation rate
$\oms$ and the disk height at the stellar radius $\hs$, and they also
have different values of the angular momentum accretion rate $j$. Note
that only two of these three variables are independent, so we can
specify a unique solution by choosing the values of two of them. We
plotted the locations of these solutions in the $\oms - \hs$ plane as
lines of constant $j$. We found that solutions with high angular
momentum accretion rates $j \sim 1$ formed Z-shaped tracks which fell
in the low-$\oms$ or the low-$\hs$ regions of the plane, as
illustrated in Figure 2 for solutions with $j=0.9$, $\md = 10^{-4.15}
\msyr$, $\ms = 0.5 \msun$, $\rs = 3 \xec$, and $\alpha = 0.01$. We
were also able to find solutions with small or negative angular
momentum accretion rates, $j ~\sles~ 0$. These solutions have moderate
values of both $\oms$ and $\hs$, as shown by the $j=0$ and $j=-1$
tracks in Figure 2. How do the spectra and line profiles produced by
these two types of solutions compare to the data?

We begin by considering high-$j$ solutions. Figure 3 shows a set of
solutions with $j=0.9$ which sit on the Z-shaped track in Figure
2. They have different values of $\oms$ and $\hs$. The low-$\hs$
solutions have a narrow boundary layer, as evidenced by the pronounced
peaks in both $\O$ and $T_{eff}$ close to $\rs$. The high-$\hs$
solutions have much broader boundary layers, and much flatter $\O$ and
$\te$ profiles. These are reflected in very different line profiles
and spectra; the high-$\hs$ solutions have strongly centrally peaked
narrow profiles as a result of their small rotational velocities
throughout the wide boundary layer. As $\hs$ decreases, the rotational
velocity in the inner disk increases, and the boundary layer gets
smaller; thus the profiles get wider, and the double-peaked disk
component becomes more prominent, forming a triple-peaked profile at
$\hs = 0.9 \xec$ and finally a double-peaked profile at $\hs = 0.8
\xec$. The spectra produced by these solutions also vary substantially
with $\hs$; as $\hs$ decreases, the smaller boundary layer reaches
higher peak effective temperatures and produces bluer spectra. None of
the high-$j$ solutions agrees with the observed spectra and line
profiles of FU Orionis systems. Only the low-$\hs$ solutions produce
reasonably double-peaked profiles, but their spectra are too blue. The
$\hs = 0.9 \xec$ solution peaks around $0.5 \mic$; the $\hs = 0.8
\xec$ solution actually peaks in the red, but is still fairly bright
in the blue and ultraviolet, unlike the observed systems.

Note that the $\hs = 0.8 \xec$ solution has both a lower luminosity
and redder peak wavelength than the $\hs = 0.9 \xec$ solution,
reversing the trend of the other solutions. This points to a more
serious problem, which was discussed in Paper I: in the low-$\hs$
solutions, much of the boundary layer is located at $R < \rs$. This
can be seen by noting that $d\O/dR$ is still quite large at $\rs$, so
that $\O$ must continue to drop inside $\rs$. Some portion of the
boundary layer flux is thus not being included in the spectrum.

In order to avoid these problems, we need to specify an additional
boundary condition. The need for an additional condition is also clear
from the fact that we have needed to specify values for two of the
three variables $\oms$, $\hs$, and $j$. In nature, one expects that
the stellar rotation rate $\oms$ is the only true variable, and that
the disk height at the stellar surface $\hs$ and the angular momentum
accretion rate $j$ will be determined by the accretion flow. An
additional condition will allow us to specify only $\oms$ in our
solutions, and the values of $\hs$ and $j$ will follow. As discussed
in Paper I, we have selected a somewhat arbitrary condition on the
radial velocity at $\rs$, $\vs = -1000 \cm \pers$. In practice, this
keeps the value of $\hs$ nearly constant with varying $\oms$. As we
showed in Paper I, $j$ drops fairly rapidly as $\oms$ increases, and
becomes negative when $\oms$ reaches 20 to 40\% of the breakup stellar
rotation rate $\omks$.

With this additional condition, we are ready to examine spectra and
line profiles for low-$j$ solutions. Figure 4 shows five solutions
with different values of $j = 0.94$, 0.9, 0.84, 0.6, 0, which lie
along the track with $\vs = -1000 \cm \pers$. The stellar rotation
rate $\oms$ increases as $j$ drops; $j = 0$ at $\oms = 1.55 \xfs
\simeq 0.3~\omks$. The high-$j$ solutions have $\O$ reaching a maximum
and then dropping down to $\oms$, while in the low-$j$ solutions, $\O$
continues to increase all the way in to $\rs$. The high-$j$ solutions
also have a more pronounced peak in $\te$, while the low-$j$ solutions
have lower effective temperatures in the boundary layer, but higher in
the disk. The high-$j$ solutions produce centrally-peaked line
profiles similar to the ones in Figs. 1 and 3. On the other hand, the
$j=0.6$ and $j=0$ solutions produce double-peaked profiles.  The
low-$j$ solutions also have redder spectra. The $j=0$ spectrum is more
luminous than the others, but it peaks in the red, and drops off
rapidly in the blue and ultraviolet due to the lack of a high-$\te$
boundary layer region. Thus, we find that low-$j$ solutions match the
observations better than high-$j$ solutions. None of the high-$j$
solutions have spectra and line profiles which agree with
observations; they either produce centrally-peaked line profiles, or
spectra which are bluer than the observed ones. The $j=0$ solution,
with moderate values of both $\oms$ and $\hs$, produces spectra and
line profiles which are in much better qualitative agreement with the
observations. Note that this conclusion is based on solutions
calculated with $\alpha = 0.01$; we discuss solutions for other values
of $\alpha$ below.

Why do low-$j$ solutions fit the observations better? Some insight
into this can be gained by noting that KHH were able to find good
agreement with spectra and line profiles of the same systems, using
standard thin disk models with the usual run of effective temperatures
and rotational velocities. This leads one to wonder why both types of
disk models, with or without boundary layers, can be successful in
reproducing the observed characteristics of FU Orionis systems. The
answer lies in the fact that the boundary layer solutions which best
fit the observations, our low-$j$ solutions, in some sense lack a
boundary layer. Conversely, more conventional high-$j$ solutions
clearly do not agree with the observations.

The low-$j$ solutions do not show the rapid drop in the angular
velocity between the disk and the star which is usually associated
with the boundary layer region. Instead, $\O$ increases slowly inward,
so that the rotational velocity in the innermost region of the disk
stays nearly constant over a wide range of radius. This produces
double-peaked line profiles like the ones observed in FU Orionis
systems. Also, $\oms$ is moderately large, generally around 20\% of
the breakup stellar rotation rate $\omks$, so that there is no part of
the inner disk which has very low rotational velocity. This inner
portion of the disk produces most of the optical flux, so any region
of it which has very low rotational velocity tends to produce a peak
in the center of the line profile, eliminating the double-peaked
structure.

A second important characteristic of the low-$j$ solutions is that
they lack the pronounced peak in the effective temperature which is
characteristic of conventional high-$j$ boundary layer solutions. In
high-$j$ solutions, this peak is produced by the rapid release of
energy which accompanies the rapid drop in $\O$. This energy is
released at fairly high effective temperatures, which produces blue
and ultraviolet fluxes substantially in excess of those observed. For
example, the high-$j$, low-$\hs$ solutions in Fig. 3 reach effective
temperatures of $\sim$ 10,000 K, whereas low-$j$ solutions for the
same parameters only reach about 7,500 K. At the same time, low-$j$
solutions are fairly luminous, as illustrated by Fig. 4. Although
these solutions are cooler than high-$j$ solutions in the innermost
region, they are hotter in the remainder of the disk. This is a result
of the different temperature distribution of a disk with a different
value of $j$. For example, if we adopt the standard thin disk
assumptions that $\O = \ok$ and that the flux radiated from the disk
surface is just the energy per unit area released by viscous
dissipation, then the effective temperature of the disk is given by
$$\te = \left({3 G \md \ms \over 8 \pi \sigma R^3} \right)^{1/4}
\left[1-j\left({\rs \over R} \right)^{1/2} \right]^{1/4}.$$ For $j=1$,
we recover the usual expression for $\te$, but for $j=0$, the factor
in square brackets is eliminated, producing higher effective
temperatures. Note that this expression only applies when the thin
disk assumptions listed above are satisfied, so that it will not apply
to the innermost portion of the disk, where $\O$ deviates strongly
from Keplerian, and radial energy transport is important. However, at
larger radii, the disk comes closer to satisfying these conditions,
and $j=0$ solutions should produce higher effective temperatures than
high-$j$ solutions. This is confirmed by Fig. 3, where the $j=0$
solution is cooler than the high-$j$ solutions at $R = \rs$, but
hotter at $R = 3 \rs$.

Solutions with $j \simeq 0$ are also ``equilibrium'' solutions which
allow the star to accrete without spinning up or down (see Paper I and
\S 5 for further discussion of this point). For these reasons, we
concentrate on $j=0$ solutions in the remainder of this paper.

\subsection{Viscosity Parameter $\alpha$}

One additional parameter that must be specified in our model is the
viscosity parameter $\alpha$. Most estimates of $\alpha$ are based
largely on the assumption that the outbursts observed in dwarf novae
proceed on the viscous timescale. This produces fairly large values of
$\alpha \sim 0.1-1$; however, application of the same idea to FU
Orionis outbursts suggests that much smaller values of $\alpha$ are
required to reproduce the long outbursts observed in these
systems. Bell \& Lin (1994) found that disk instability models for FU
Orionis systems required values of $\alpha \sim 10^{-4}$ where
hydrogen is neutral and $\alpha \sim 10^{-3}$ where hydrogen is
ionized. However, such small values of $\alpha$ lead to problems with
evolutionary timescales and gravitational instability in the disk
(Bell \et 1995).

In this work, we have adopted a compromise value of $\alpha = 10^{-2}$
for most of our models. In order to explore the effects of varying
$\alpha$ in our models, we have calculated a sequence of solutions
with $\log \alpha = -1.0, -1.5, -2.0, -2.5, -3.0$. These solutions all
have the same values of $\md = 10^{-4.15} \msyr$, $\ms = 0.5 \msun$,
$\rs = 3 \xec$, and $j=0$, and we have varied the value of $\vs$ such
that $\vs = -1000 (\alpha / 0.01) \cm \pers$. These solutions are
shown in Figure 5. 

The larger values of $\alpha$ correspond to thinner disks, which
produce faster stellar rotation rates and higher effective
temperatures in the inner disk. If we were to continue to increase
$\alpha$ (perhaps to unreasonably large values $\alpha > 1$),
eventually we would reach a nearly Keplerian, thin disk solution. Such
a solution would have the star rotating near breakup, and would
resemble the solutions found by KHH. The smallest value, $\alpha =
10^{-3}$, produces a solution in which the rotation rate is very small
throughout the inner disk. This results in a very narrow,
centrally-peaked line profile. Larger values of $\alpha$ produce
progressively wider double-peaked profiles. There is also a clear
trend in the spectra as $\alpha$ changes; larger values of $\alpha$
produce bluer spectra and higher total luminosities. The variation in
luminosity is due to the advection of a larger fraction of the
accretion luminosity into the star in the small-$\alpha$ solutions,
where the disk has become quite thick.

\section{Comparison to Observed Systems}

We now use our disk and boundary layer solutions to fit the spectra
and line profiles of the two best-observed FU Orionis systems: FU
Orionis and V1057 Cygni. Both have resolved line profiles available in
both the optical and near-infrared, along with broad-band photometry
which illustrates the shape of the spectral energy distribution from
3600 \ang out to 10 $\mic$ and beyond.  These were two of the original
three systems discussed by Herbig (1977) in his seminal paper. The
third system, V1515 Cygni, has very narrow, unresolved line profiles
which suggest that it is probably viewed pole-on, making it difficult
to constrain models for this system. Another well-observed system
which is a probable member of the FU Orionis class is Z CMa, which has
very broad line profiles. Unfortunately, this system has a binary
companion which greatly alters the shape of the infrared spectrum. The
other candidate systems have some of these data available, but not all
of them, and for many of them their identification as FU Orionis
systems is still somewhat uncertain (see Hartmann \& Kenyon 1996 for a
discussion of these objects).

\subsection{Observations}

Most of the observations of FU Orionis and V1057 Cygni used in this
work are the ones which were used by KHH in their comparison to
Keplerian thin disk models. These consist of high-resolution spectra,
which are used to derive line profiles, and photometry to show the
broad-band spectral energy distributions of these objects. The
high-resolution spectra include echelle spectra of a 50-\ang region
around 6170 \ang, and Fourier transform spectra of the 2.2 $\mic$
region. We have used a more recent 2.2 $\mic$ spectrum of V1057 Cyg
with improved resolution of $\sim 10 \kms$. Unlike the earlier
spectrum, the cross-correlation produced from this spectrum is clearly
double-peaked. As described by KHH, the photometric data include
low-resolution optical spectrophotometry which has been binned into
the standard photometric bands, along with standard optical and
infrared photometry. The photometric data have been dereddened using
the extinction law given by Savage \& Mathis (1979) and assuming $A_V
= 2.2$ mag for FU Orionis and $A_V = 3.5$ mag for V1057 Cygni.

\subsection{Fitting Procedure}

Based on the results discussed in \S 3, in attempting to fit the
observations, we have fixed certain parameters: since low-$j$
solutions seem to fit the general characteristics of the observations,
we have fixed $j=0$. We also have fixed the boundary condition on $\vs
= -1000 \cm \pers$. This together with the condition on $j$ fixes the
position of the solution in the $\oms - \hs$ plane (see Fig. 2). We
also fix the value of $\alpha $ at $10^{-2}$. This leaves the standard
accretion parameters $\md$, $\ms$, and $\rs$, which we vary in order
to fit the observations. As discussed in \S 3.1, we can use these
three parameters to calculate $L_{acc} = G \ms \md / \rs$, $T_{max} =
0.287 (G \ms \md / \sigma \rs^3)^{1/4}$, and $v_K(\rs) = (G \ms /
\rs)^{1/2}$, which relate more directly to the observations which we
are trying to match. These parameters give us a good sense of how
changes in $\md$, $\ms$, and $\rs$ will affect our spectra and line
profiles. For instance, if we want to increase the luminosity of a
particular solution without changing the temperature or rotational
velocity, the dependences given above suggest that we need to keep the
ratios $\md \ms / \rs^3$ and $\ms / \rs$ constant. If we increase
$\rs$, we can increase $\ms$ by the same factor and increase $\md$ as
$\rs^2$. This will increase $L_{acc}$ as $\rs^2$ while keeping
$T_{max}$ and $v_K(\rs)$ constant. 

There are several other parameters which enter into our fits: the
inclination angle, distance, and extinction of the individual FU
Orionis systems. The inclination angle $i$ and the distance $d$ affect
the derived luminosity of the FU Orionis systems, since we convert the
luminosity of our solutions to flux using the relation $F_{\lambda} =
L_{\lambda} \cos i / 2 \pi d^2$, and then compare these fluxes to the
observed ones. The inclination angle also affects the linewidths
derived from our solutions, since the linewidth is set by the
line-of-sight velocity of the disk material $v_{los} = v_{rot} \sin
i$. The extinction $A_V$ affects both the shape and the luminosity of
the spectra, since the observed photometric points are dereddened
before they are used for comparison with the solutions. We adopt
values published elsewhere for the distance (Hartmann \& Kenyon 1996)
and extinction (KHH), and leave these fixed. We vary the inclination
as part of our fitting procedure.

Our fits to the observations are all approximate, and are done by
eye. In comparing the spectra produced by our solutions with
photometric data, we weight some of the data points less heavily than
others. At wavelengths longer than 10 $\mic$, the data frequently show
that the objects are much brighter than our models would predict. This
is generally attributed to the presence of reprocessing of some of the
disk luminosity by the outer disk or by a dusty envelope surrounding
the disk (KHH; Kenyon \& Hartmann 1991). Thus we confine our
comparison with the photometric data to wavelengths shorter than $10
\mic$. At the blue end of the spectrum, the U-band point at 3600 \ang
is generally substantially fainter than the B-band point at 4400
\ang. This rapid drop in the spectral energy distribution is probably
due largely to the Balmer jump, and it is therefore difficult to
reproduce it with a blackbody spectrum. Our spectra produced using a
library of stellar spectra tend to be closer to the observed point. A
number of other factors could also contribute to the rapid dropoff
from B to U; the extinction increases dramatically as one moves to
shorter wavelengths in this portion of the spectrum, and so error in
the dereddening applied to the observed spectrum will have their
greatest effect here. Also, inclination effects tend to depress the
blue end of the spectrum.

When we compare our line profiles to the data, we do not compare them
directly to individual line profiles, which tend to be noisy and
frequently blended with other nearby lines, but rather to the shape of
the cross-correlation function. The cross-correlation function
represents the mean line shape in the spectral region for which it is
derived. We have assigned the zero point of our model line profiles to
the zero point of the cross-correlations, and normalized the two by
their peak heights. We quantitatively compare the overall shape of our
line profile to the cross-correlation by determining the velocity
widths of the two at half-maximum and at full-maximum. We also compare
the ratio of the optical and infrared line widths produced by our
solutions to the ratio seen in the data.

\subsection{Fits to Individual Systems}

\subsubsection{FU Orionis}

This object has broad line profiles, with full-width at half-maximum
of $100 \kms$ at 6170 \ang and $74 \kms$ at 2.2 $\mic$. It is also
quite bright, with a peak value of $\log \lambda F_{\lambda} \simeq
-7.5$ in the V-band, assuming $A_V = 2.2$. Adopting a distance of 500
pc, we find good fits to the line profiles and spectrum from a
solution with $\md = 2 \times 10^{-4} \msyr$, $\ms = 0.7 \msun$, $\rs
= 4 \xec$, and $i = 60^\circ$. As mentioned above, the solution has
$j=0$ and $\alpha = 0.01$. It also has $\vs = -1000 \cm \pers$, giving
a stellar rotation rate of $8.29 \xss$ and a disk height at the stellar
radius $\hs = 1.57 \xec$. These values correspond to a stellar
rotation period of 8.76 days and $\hs/\rs = 0.39$.

The angular velocity $\O$ and rotational velocity $v_{rot} = \O R$ are
shown in Figure 6a. The rotational velocity at the stellar surface is
only $33 \kms$, and the maximum rotational velocity is $47 \kms$ at $R
\simeq 8.75 \xec \simeq 2.19 \rs$. Figure 6b shows the effective
temperature $\te$, which reaches a maximum value of 8212 K at $R
\simeq 4.6 \xec = 1.15 \rs$.  Figure 6c shows the line profiles
derived from this disk model at 6170 \ang and 2.2 $\mic$,
respectively. Both profiles are double-peaked, like the
cross-correlations, and we find FWHM values of 98 and $72 \kms$ at the
two wavelengths. Note that the cross-correlation peaks are rather
asymmetric; this is probably due to the presence of a mass loss in a
wind, as discussed by KHH.  Figure 6d shows the fits of blackbody and
stellar composite spectra to the photometry of FU Orionis. Both
spectra fit the data points quite well from the B band to 10 $\mic$,
deviating from most data points by 10\% or less. Both are too bright
in the U band, although the stellar composite spectrum comes
substantially closer to the observed point. The luminosity emitted by
the boundary layer and disk is $L \simeq 1.83 \times 10^{36} \ergs
\simeq 470 \lsun$.

If this were a standard Keplerian thin disk, the disk parameters
$\md$, $\ms$, and $\rs$ for this solution would give an accretion
luminosity $L_{acc} \simeq 2.95 \times 10^{36} \ergs \simeq 750
\lsun$, a maximum disk temperature $T_{max} \simeq 6850$ K, and a
Keplerian rotational velocity $v_K(\rs) \simeq 150 \kms$. These are
substantially different from the actual luminosity, peak effective
temperature, and peak rotational velocity found in our solution. These
differences indicate the importance of including the boundary layer
region in our disk model.

\subsubsection{V1057 Cygni}

This system has narrower line profiles than FU Orionis, with FWHM
values of $57 \kms$ at 6170 \ang and $47 \kms$ at 2.2 $\mic$. It also
appears to be somewhat fainter; after dereddening with $A_V = 3.5$, it
has a peak $\log \lambda F_{\lambda} \simeq -7.8$, and the peak occurs
in the R band, indicating that this system is slightly cooler than FU
Orionis. We adopt a distance of 600 pc. 

We find good fits to the spectrum and line profiles of V1057 Cygni
from a disk and boundary layer solution with $\md = 10^{-4} \msyr$,
$\ms = 0.5 \msun$, and $\rs = 3.5 \xec$. This solution is less
luminous and slightly cooler than the solution for FU Orionis
described above, but in most other respects the two solutions are
quite similar. Like the solution for FU Orionis, this solution has
$j=0$, $\alpha = 10^{-2}$, and $\vs = -1000 \cm \pers$. This gives a
stellar rotation rate $\oms = 8.88 \xss$, which corresponds to a
rotation period of 8.19 days. The disk height at $\rs$ is $\hs = 1.37
\xec$, so $\hs/\rs = 0.39$.

The angular and rotational velocities of this solution are shown in
Figure 7a. The rotational velocity is $31 \kms$ at the stellar
surface, and reaches a peak value of $49 \kms$ at $R = 8.49 \xec =
2.43 \rs$. The effective temperature, shown in Figure 7b, reaches a
maximum value of 7111 K at $R \simeq 4 \xec = 1.14 \rs$.  Figure 7c
shows the line profiles at 6170 \ang and 2.2 $\mic$ for this
solution. The profiles have FWHM values of 54 and $44 \kms$,
respectively, compared to the observed values of 57 and $47 \kms$. The
ratio of the optical to infrared linewidth is quite close to the
observed ratio, where the KHH thin disk model gave infrared linewidths
which were about 25\% smaller than those observed. The separation of
the two peaks is 26 and $14 \kms$ in our disk solution, and 33 and $19
\kms$ in the data.  The fits of our blackbody and stellar composite
spectra to the photometric data for V1057 Cygni are shown in Figure
7d. The blackbody spectrum fits quite well in both the optical and
near-infrared regions; it is too bright in the U-band, as
expected. The data show a drop in the I-band and an excess at 3.5 and
4.8 $\mic$; these points would be difficult to fit with any smooth
curve which also fits the optical and near-infrared points. The
stellar composite spectrum comes much closer to the U-band point but
is slightly fainter than the data in the optical.

\subsection{Comparison to Thin Disk Models}

It is interesting to compare the disk parameters $\md$, $\ms$, and
$\rs$ of our best-fit models for FU Orionis and V1057 Cygni to those
found by KHH using thin disk models. KHH list $\ms$ and $\rs$ for each
system as a function of the inclination angle $i$; these values of
$\ms$ and $\rs$ keep the projected rotational velocity at the stellar
surface $(G \ms / \rs)^{1/2} \sin i$ constant at $93 \kms$ for FU
Orionis and $42.6 \kms$ for V1057 Cygni. KHH also keep the maximum
disk temperature constant at 7200 K in FU Orionis and 6590 K in V1057
Cygni, so $\md$ also varies with inclination. Our models use an
inclination of $60^\circ$ for FU Orionis and $30^\circ$ for V1057
Cygni. For these inclinations, KHH found $\ms = 0.34 \msun$, $\rs =
5.47 \rsun$, and $\md = 4.36 \times 10^{-4} \msyr$ for FU Orionis, and
$\ms = 0.15 \msun$, $\rs = 4.02 \rsun$, and $\md = 2.75 \times 10^{-4}
\msyr$ for V1057 Cygni.

Our best-fit models give $\ms = 0.7 \msun$, $\rs = 5.75 \rsun$, and
$\md = 2 \times 10^{-4} \msyr$ for FU Orionis, and $\ms = 0.5 \msun$,
$\rs = 5.03 \rsun$, and $\md = 10^{-4} \msyr$ for V1057 Cygni. Thus we
find stellar masses which are a factor of 2--3 larger, mass accretion
rates which are a factor of 2--3 smaller, and slightly larger stellar
radii. These differences are due to the fact that our models include
the boundary layer region. Thus, rotational velocities are
substantially below Keplerian in the inner disk, requiring larger
stellar masses to produce the observed linewidths. To compensate for
this increased stellar mass, the mass accretion rate must decrease by
a similar factor in order to produce similar luminosities and
temperatures. Note that the KHH models only included the disk
luminosity, which is half of the total accretion luminosity in the
standard thin disk model. Our models include both the disk and
boundary layer luminosities, but they do not radiate the full
accretion luminosity, as mentioned in \S 4.3 and discussed below.
Finally, the presence of the boundary layer increases the inner disk
temperatures, and so a slightly larger stellar radius is required to
keep the spectra from becoming too blue.

\section{Discussion}

\subsection{Standard Disk Solutions vs. Disk and Boundary Layer
Solutions}

We have found disk and boundary layer solutions with significant
departures from Keplerian rotation that can reproduce the key spectral
features of FU Orionis and V1057 Cygni.  However, observations of
these objects were originally modelled with reasonable success using
standard Keplerian disk theory (Hartmann \& Kenyon 1985; KHH).  Here
we discuss why it is difficult to compare the relative merits of the
two types of solutions on the basis of observations alone.  We then
outline why we think the boundary layer solutions are nevertheless an
important improvement in treating FU Orionis accretion physics.

Both standard thin disk solutions and disk and boundary layer
solutions agree quite well with observations. Thus, comparisons with
observations do not provide a definitive test of the relative merits
of the two types of solutions for FU Orionis objects.  The disk and
boundary layer solutions appear to match the differential rotation
with wavelength somewhat better than the thin disk solutions of KHH;
the thermal pressure support in the inner regions of the disk tends to
reduce the ratio of the optical line widths to the near-infrared line
widths, as observed.  However, as pointed out by KHH, there are
significant uncertainties in strengths of the infrared CO lines in the
outer disk due to the possibility of dust formation, and with
plausible models of dust contributions a thin disk in Keplerian
rotation can match the observed differential rotation with wavelength.

Similarly, differences in how well the two types of solutions match
the observed spectral energy distributions, or small differences in
the inferred masses, radii, and accretion rates, are not significant
in view of the many uncertainties involved. Specifically, there are
uncertainties in the inclination angles $i$, distances $d$, and
extinctions $A_V$ of FU Orionis and V1057 Cygni. We have used fairly
standard inclinations of $i = 60^\circ$ for FU Orionis and $i =
30^\circ$ for V1057 Cygni. These inclinations allow solutions with
similar rotational velocities to produce the rather different
linewidths seen in these two systems. In fact, the ratio $\sin
30^\circ / \sin 60^\circ \simeq 0.58$ is almost identical to the ratio
of the 6170 \ang linewidths (57 and $100 \kms$) in these objects.
Nonetheless, other inclinations, in combination with different choices
of the other solution parameters, might produce solutions which match
the data. The distances to FU Orionis and V1057 Cygni are probably
only accurate to 10--20\%, so that the luminosities inferred for these
systems could be in error by 20--40\%. Finally, the extinctions of
these systems are not well known, and could represent an additional
source of error in both the luminosities and the shape of the spectra.
All of these considerations limit our ability to distinguish between
different disk models.

Nonetheless, we feel that the disk and boundary layer model presented
in this paper offers significant advantages over the standard thin
disk model. First, it avoids some uncomfortable assumptions which are
implicit in the thin disk model.  The two major observational
constraints on models of the inner disk in FU Orionis systems are the
apparent lack of hot boundary layer emission and the lack of
slowly-rotating material in the observed line profiles. The absence of
these features gives the impression that the boundary layer is not
present is FU Orionis systems.  The standard thin disk model (Hartmann
\& Kenyon 1985; KHH) simply did not include any boundary layer
region. By omitting the boundary layer, this model made the implicit
assumption that the accreting star is rotating at breakup speed. Rapid
stellar rotation seems unlikely in view of the generally slow
rotations of T Tauri stars, plus the recognition that FU Orionis and
V1057 Cygni have not accreted enough angular momentum during their
outbursts to spin up the entire central star as a solid body (Hartmann
\& Kenyon 1996). This leaves open the possibility that only the outer
layers are spun up to Keplerian angular velocities. A second implicit
assumption of the early model is that the disk can be Keplerian all
the way in to the stellar surface, which implies that pressure
gradients in the inner disk are insignificant. This also seems
unlikely; at the high mass accretion rates found in FU Orionis
systems, radial pressure gradients will support the accreting material
and produce sub-Keplerian angular velocities unless $\alpha$ is very
large.

Our disk and boundary layer model avoids these assumptions. We have
found that by using the slim disk model, we can include the boundary
layer region in our calculations of FU Orionis disks and still find
solutions which agree with the observations. These solutions
demonstrate explicitly what the thin disk model implicitly assumed: FU
Orionis disks do not have a standard accretion disk boundary layer
where the angular velocity of the accreting material drops rapidly
over a short radial distance.  Unless $\alpha$ is very large, the high
accretion rates of FU Orionis objects require quite optically thick
disks which are very hot in their interiors, resulting in signficant
thermal pressure support.  Thus the narrow, hot boundary layer
typically expected in pre-main-sequence stars (Lynden-Bell \& Pringle
1974) automatically disappears, for a wide range of stellar rotation
rates.  Instead, the ``boundary layer'' is a broad region where $\O$
changes gradually; in our best-fitting solutions, $\O$ increases
gradually and becomes nearly constant as the accreting material
approaches the star. As a result, like the early thin disk solutions,
our solutions lack the high-temperature, slowly-rotating region
expected from a standard boundary layer. But unlike the early
solutions, they do not require that the star be rotating at breakup or
that $\alpha$ be very large.  In fact, for $\alpha = 10^{-2}$, our
best solutions have the star rotating at only $\sim 20-25\%$ of
breakup speed. Fig. 5 shows that as $\alpha$ increases, $\oms$
increases, reaching $\sim 50\%$ of breakup speed at $\alpha =
0.1$. This suggests that the Keplerian thin disk solutions of Hartmann
\& Kenyon (1985) and KHH represent a special, extreme case of the
solutions found in this paper, where the values of both $\alpha$ and
$\oms$ are very large.

In addition to eliminating the assumptions of very large values of
$\alpha$ and $\oms$ which are implicit in the thin disk model, our
disk and boundary layer model can offer insight into new areas of FU
Orionis accretion physics. The model explicitly includes the effects
of the rotation rate of the star on the boundary layer region and on
the angular momentum accretion rate.  Since radial advection of energy
is included in the slim disk equations, our solutions directly give
the rate at which energy is carried into the accreting star. These
issues can only be studied when the boundary layer region and the
additional physics of the slim disk equations are included in the
model. They have important consequences for the evolution of
pre-main-sequence stars, which will be discussed in more detail in the
following sections.
 
\subsection{Low-$j$ Solutions: Implications for Spin Evolution of
Pre-Main-Sequence Stars}

Our best-fit solutions for FU Orionis and V1057 Cygni are equilibrium
solutions with $j=0$, i.e. the central star is neither gaining nor
losing angular momentum, and have stellar rotation rates $\oms \simeq
0.225 \omks$ which correspond to rotation periods of 8--9 days. These
periods are comparable to those observed in T Tauri stars (Bouvier \et
1993, 1995). These low-$j$ solutions have important implications for
the spin evolution of pre-main-sequence stars, because they support
the scenario proposed in Paper I, where FU Orionis outbursts regulate
the rotation rates of T Tauri stars. In Paper I we showed that the
angular momentum accretion rate $j$ drops rapidly as the stellar
rotation rate $\oms$ increases, and reaches $j=0$ when $\oms \simeq
0.2 - 0.4 \omks$ for FU Orionis parameters. For smaller values of
$\oms$, $j > 0$ and the star spins up, while for larger values of
$\oms$, $j < 0$ and the star spins down. Thus, FU Orionis outbursts
will move $\oms$ toward an equilibrium value where $j \simeq 0$. If FU
Orionis outbursts dominate mass accretion onto T Tauri stars, as seems
likely from event statistics, then they should also dominate angular
momentum accretion and control the spin evolution of these stars. When
the T Tauri star is between outbursts, it may spin up or down, but
during each outburst it will return to the equilibrium rotation
rate. This scenario therefore predicts that FU Orionis systems should
attain a low-$j$ equilibrium state during outbursts, and we find that
low-$j$ solutions fit the observations.

We are unable to make more precise predictions of the spin rates of T
Tauri stars because of uncertainties in the structure of the central
star undergoing FU Orionis accretion.  As discussed in \S 5.4 below,
it is likely that the advection of large amounts of thermal energy
through the inner disk will cause the central star to expand from its
equilibrium state.  After an outburst ceases, we expect that the star
will contract to a smaller radius, and will therefore spin up somewhat
in the absence of any angular momentum loss.  Note that the disk may
still remove large amounts of angular momentum from the star and spin
the star down substantially during the course of the outburst;
however, at the end of the outburst, we expect that there will always
be some spinup due to the contraction of the star. Thus the
equilibrium rotation rates during the FU Orionis phase should
correspond to faster T Tauri rotation rates. The amount of spinup will
depend upon the moment of inertia of the expanded layers of the star,
which we are not able to estimate at present.

Observations of T Tauri stars suggest that classical T Tauri stars
which have accretion disks rotate more slowly than weak-line T Tauri
stars which lack disks (Bouvier \et 1993, 1995; Edwards \et 1993;
Eaton, Herbst, \& Hillenbrand 1995; Choi \& Herbst 1996). Our results,
combined with those of Paper I, suggest a new way in which the disk
can regulate the rotation rate of an T Tauri star.  Previously
suggested methods of regulating the stellar rotation rate have relied
upon the interaction between the stellar magnetic field and the disk
(K\"onigl 1991; Cameron \& Campbell 1993; Hartmann 1994; Shu \et
1994). Our method does not depend on the stellar magnetic field, which
is presumably not strong enough to disrupt the disk during FU Orionis
outbursts. However, it does require that outbursts occur, and that
they dominate the angular momentum accretion onto the star.

\subsection{Luminosities}

We noted in \S 4.3 that our solutions for FU Orionis and V1057 Cygni
produce luminosities which are substantially smaller than the
accretion luminosity. For FU Orionis, we found that our solution gave
$L = 470 \lsun$, whereas $\lacc \simeq 750 \lsun$, so that $L \simeq
0.63 \lacc$. For our V1057 Cygni solution, $L \simeq 195 \lsun$ and
$\lacc \simeq 310 \lsun$, so we again have $L \simeq 0.63 \lacc$. The
remaining luminosity goes into heating the disk material. At the
stellar surface, the midplane temperature of the disk in our solutions
becomes quite large: $T_c(\rs) \simeq 2.5 \times 10^5$ K in the FU
Orionis solution, and $2 \times 10^5$ K in the V1057 Cygni solution.
This means that the disk material carries energy into the star at the
rate $\md c_P T_c(\rs) \simeq 280 \lsun$ for FU Orionis and $115
\lsun$ for V1057 Cygni. The high temperature of the disk material is
due to the large vertical optical depth of the disk and boundary layer
in these solutions.  If the accreting stars in FU Orionis objects are
typical pre-main-sequence stars, they should have luminosities $\sim 1
- 10 \lsun$, so this advected energy represents a major perturbation
to the star.

Another interesting feature of these solutions is that only a fraction
of the luminosity comes from energy liberated by viscous
dissipation. Popham \& Narayan (1995) derived an expression for the 
total viscous dissipation rate of the disk and boundary layer
$$
L_{diss} = {G \ms \dot M \over \rs} \left( 1 - j {\oms \over \omks} +
{1 \over 2} {\oms^2 \over \omk2s} \right) 
+ \dot M {\int {dP \over \rho}},
$$
where we have left out several small terms. The first term in this
expression represents the gravitational potential energy and the
rotational kinetic energy transferred between the star and the
disk. For $j=0$, the disk is removing a small amount of rotational
kinetic energy from the star. Both of our solutions have $\oms / \omks
= 0.225$, so the first term is just $1.025 L_{acc}$. In a thin disk,
the second term is very small, but in our FU Orionis solutions, this
term becomes quite large due to the importance of the radial pressure
gradient in supporting the accreting material against gravity. In both
solutions, we find $\md {\int dP / \rho} = -0.74 L_{acc}$, so that the
viscous dissipation rate is only $L_{diss} = 0.285 L_{acc}$.
Nonetheless, we know that both solutions radiate $L \simeq 0.63
L_{acc}$; the remaining $\sim 0.35 \lacc$ comes from the entropy term
in the energy equation
$$
\nu \Sigma \left ( R{d\Omega \over dR}\right ) ^2 - F_V -\Sigma v_R
T_c {dS \over  dR} - {1 \over R}{d \over dR}(RHF_R) = 0.
$$
The four terms in this equation represent the viscous dissipation, the
vertical flux from the disk surface, the advected entropy, and the
divergence of the radial flux. If we integrate this equation over the
surface of the disk, we have
$$
L_{diss} - L + \md {\int T dS} - 4 \pi [R_{out} H_{out} F_{R,out} -
\rs \hs F_{R,*}] = 0.
$$
Our boundary conditions assume that the radial fluxes at the inner and
outer edges of the disk are small; at the inner edge we assume that
the flux entering the disk is $\sigma T_*^4$, where we assume $T_* =
5000$ K, so $4 \pi \rs \hs F_{R,*} \sim 0.01-0.02 \lacc$, and at the
outer edge the radial flux is insignificant. If we neglect these
radial flux terms, we can use $T dS = dU - P d \rho / \rho^2$ and the
expression for $L_{diss}$ given above to write
$$
L = L_{acc} \left( 1 - j {\oms \over \omks} + {1 \over 2} {\oms^2
\over \omk2s} \right) + \md {\int {dP \over \rho}} + \md {\int dU} -
\md {\int {P \over \rho^2} d\rho}
$$
$$
\simeq L_{acc} \left( 1 - j {\oms \over \omks} + {1 \over 2} {\oms^2
\over \omk2s} \right) + {5 \over 2} \md {\int d \left({P \over
\rho}\right)},
$$
$$
\simeq L_{acc} \left( 1 - j {\oms \over \omks} + {1 \over 2} {\oms^2
\over \omk2s} \right) - \md c_P T_c(\rs)
$$
where the internal energy $U \simeq 3 P / 2 \rho$, $c_P T_c \simeq 5 P
/ 2 \rho$, and $T_c(\rs) \gg T_c(\ro)$. This accounts for the
difference between the actual luminosities radiated by our solutions
and the accretion luminosities for those disk parameters.

\subsection{Stellar Radii and Advected Energy}

Our best-fit solutions for FU Orionis and V1057 Cygni have stellar
radii of $5-6 \rsun$. These are substantially larger than the radii of
T Tauri stars, which are generally estimated at $1.5 - 3 \rsun$
(Bouvier \et 1995). Since FU Orionis outbursts are believed to occur
in T Tauri star accretion disks, this seems to imply that the star
expands rapidly during the outburst, probably as a result of the rapid
addition of high-temperature material (KHH; Hartmann, Cassen, \&
Kenyon 1996).

Prialnik \& Livio (1985) calculated the effects of accretion onto a
$0.2 \msun$, $0.2 \rsun$ fully-convective main-sequence star. They
found that if the accretion carries a sufficient amount of energy into
the star, it can cause the star to expand stably or unstably. For
accretion rates comparable to those of FU Orionis stars, their
calculations indicate that the expansion will proceed unstably if the
accretion energy is added to the star at a rate faster than about 10\%
of the accretion luminosity.  The rapid expansion is due to the
conversion of the normally convective star to a convectively-stable,
radiative structure.

As the discussion of the luminosities shows, our solutions add energy
to the accreting star at an enormous rate - $0.375 L_{acc} \simeq 100 -
300 \lsun$ in our solutions for FU Orionis and V1057 Cygni. This rate
of energy transfer will almost certainly produce rapid expansion of
the star. The character of this expansion is not well understood. The
accreting material comes from a disk, so it carries angular momentum
as well as energy, and is added around the star's equator. Thus the
expansion is likely to be nonspherical. More sophisticated models of
this process are needed, but the problem is clearly a difficult one.
The large energy input into the star and the resulting stellar
expansion should have important implications for the decline from FU
Orionis outbursts. The star will contract and release the energy it
gained during the outburst, which is a substantial fraction of the
total accretion energy of the outburst.  Thus we predict,
qualitatively, that at the end of the mass accretion outburst in the
disk, there may be a phase in which the central star is overly
luminous, and fades over some (rapid) time set by the Kelvin time of
the perturbed portion of the stellar envelope, similar to the original
explanation put forth by Larson (1983) for FU Orionis outbursts.

\subsection{Future work}

By including the boundary layer region and the disk structure
self-consistently, the solutions presented here represent a
substantial advance in our understanding of disks in FU Orionis
objects and the spectra and line profiles they produce. Nonetheless,
there are additional explorations which could be made within
the context of our assumptions, and improvements which could be
made to our model.

The solutions we have found do a good job of fitting the observed line
profiles and spectra of FU Orionis and V1057 Cygni. Nevertheless, this
does not mean that they are the only solutions which would fit the
observations of these systems. The number of solution parameters is
too great to permit a full exploration of the entire parameter
space. Accordingly, as described in \S 4.2, we have used the standard
parameters $\lacc$, $T_{max}$, and $v_K(\rs)$ to guide our search. We
have also kept certain parameters constant in order to simplify the
fitting procedure. We keep $j=0$ for the reasons discussed in \S 3,
where we demonstrated that high-$j$ solutions fail to produce spectra
and line profiles that agree with observations. Other values of $j$
could produce reasonable solutions; for instance, in Paper I we
proposed that FU Orionis outbursts may spin down the accreting star,
which would require solutions with negative values of $j$.
Negative-$j$ solutions are similar in all respects to $j=0$ solutions;
the boundary layer region has $\O$ increasing inward throughout and
lacks a strong peak in effective temperature. As we demonstrated in
Paper I, $j$ drops rapidly with increasing $\oms$, so that a $j=-1$
solution has $\oms$ only slightly larger than a $j=0$ solution with
the same parameters (see Fig. 2). 

We also use $\alpha = 10^{-2}$ for all of our solutions. As discussed
in \S 3.3, solutions with smaller values of $\alpha$ have thicker
disks, so that a larger fraction of the accretion luminosity is
advected into the star. Thus a smaller fraction is radiated, and the
effective temperature is lower. Another consequence of changing
$\alpha$ is that the rotational velocities change; small values of
$\alpha$ result in small rotational velocities which would produce
centrally peaked line profiles unlike those observed. This suggests
that values of $\alpha > 10^{-2}$ may produce acceptable solutions,
but those with $\alpha$ substantially below $10^{-2}$ may not.  This
is particularly interesting because Bell \& Lin (1994) and Bell \et
(1995) found that they needed $\alpha \sim 10^{-3} - 10^{-4}$ to
obtain satisfactory outbursts using the thermal instability mechanism.

Our model spectra could also be improved by treating the effects of
inclination in more detail. We have implicitly assumed that the disk
surface is flat, so that the inclination angle for all parts of the
disk surface is the same. In fact, most of our solutions have $H/R$
nearly constant with $R$, with $H/R \sim 0.4$. If we take $H(R)$ as
representing the surface of the disk, then the disk surface is
inclined at $\tan^{-1} (H/R) \sim 22^\circ$ to the disk midplane. This
means that at inclination angles greater than $i \sim 68^\circ$, a
portion of the disk surface will not be visible. Even if $i <
68^\circ$, the side of the disk closer to the observer will
effectively be seen at a larger inclination angle than the opposite
side. (note that when we refer to ``sides'' of the disk here, we are
referring not to the top and bottom surfaces of the disk, but to
regions of the same surface at different angles around the rotation
axis.) Also, one side of the disk will be heated by radiation from the
other side. We have also assumed that the disk spectrum does not vary
with inclination angle, whereas in fact limb darkening will generally
produce a decline in the blue end in the spectrum for systems viewed
at large inclination angles.

Finally, there are improvements which could be made in the physical
treatment of the disk and boundary layer.  The assumption of a steady
disk might be relaxed, although we feel this is unlikely to be a major
issue, even though we are comparing our solutions to objects
experiencing outbursts, because FU Orionis and V1057 Cygni have both
remained fairly steady over recent years (KHH; Kenyon \& Hartmann
1991), and we are only addressing the innermost regions of the
disk. Time-dependent models of boundary layers in disks around
pre-main-sequence stars (Godon 1996) also seem to agree with our
steady-state results (PNHK).  Calculations of the structure of
advection-dominated accretion disks (Narayan \& Yi 1995) have
demonstrated that the slim disk equations used in our model provide a
fairly good representation of the disk structure even for disks which
are quite vertically thick; still, a two-dimensional model for the
boundary layer and disk might provide insights into FU Orionis objects
that we are unable to make using our current model (e.g., Kley 1991),
especially when considering a more subtle matching of the disk to the
inherently two-dimensional star.  Some consideration should also be
given to understanding the effects of rapid accretion on the central
star. Presumably these include rapid expansion of the star, which
could have important effects on the boundary layer region and the
angular momentum accretion by the star. Ultimately this will have to
be a two-dimensional, time-dependent calculation as well, but perhaps
some progress can be made with calculations similar to those performed
by Prialnik \& Livio (1985) for main-sequence convective stars.

\section{Summary}

We have calculated solutions for the structure of the accretion disk
in FU Orionis objects which include a self-consistent treatment of the
boundary layer region. We have also computed the line profiles and
continuum spectra that would be observed from these solutions. We have
explored the dependence of these observable characteristics on the
solution parameters, including the mass accretion rate, the stellar
mass, radius, and rotation rate, the angular momentum accretion rate,
and the viscosity parameter.  We find that our slim disk solutions can
account for the absence of hot boundary layer emission without
requiring that the central star rotate at breakup velocity.  The
differential rotation of the innermost disk departs from Keplerian
rotation and gives marginally better agreement with observations than
a Keplerian disk, but uncertainties in line formation preclude using
the observations to make a definitive test.  We find solutions at spin
equilibrium ($j \sim 0$) for stellar angular velocities comparable to
those observed in T Tauri stars, and these solutions fit the
observations of FU Orionis and V1057 Cygni. This result supports the
proposal made in Paper I that FU Orionis outbursts play an important
role in regulating the rotation rates of T Tauri stars.

In our solutions, large amounts of thermal energy are being advected
into the central star during an FU Orionis outburst. This should cause
the star to expand, which would help to explain why the central stars
of FU Orionis objects appear to be twice as large as typical
pre-main-sequence stars.  The heating and expansion of the star should
produce an observational signature: the fading of the central star
after the end of rapid disk accretion.  Further theoretical
explorations of the effects of rapid disk accretion on
pre-main-sequence stars are needed.

\acknowledgements

RP acknowledges the support of grants NASA NAG5-2837 and NSF
AST-9423209 at the Center for Astrophysics and grants NASA NAGW-1583,
NSF AST 93-15133, and NSF PHY 91-00283 at the University of Illinois.
The research of LH and SK is supported in part by NASA grant
NAGW-2306.

\newpage

\figcaption{Three boundary layer solutions with different values of
$\md$ and $\rs$; the solutions have $\md = 10^{-4.3}, 10^{-4.15},
10^{-4.0} \msyr$, and $\rs = 2.25, 3, 4 \xec$, respectively, and are
labeled by their values of $\rs$ in units of $10^{11} \cm$. These are
the choices of $\md$ and $\rs$ that were used in Paper I; note that
they keep the accretion luminosity approximately constant. All three
solutions have high angular momentum accretion rates, with $j = 1.0$,
0.9, 0.8, respectively, and all three have $\ms = 0.5 \msun$,
$\alpha = 10^{-2}$, and $\vs = -1000 \cm \pers$. The four panels show
(a) the angular velocity $\O$; the Keplerian angular velocity for
these parameters is shown by the dashed line; (b) the effective
temperature $\te$; (c) line profiles at 6170 \ang, assuming an
inclination $i = 60^\circ$ and instrumental broadening of $12.5 \kms$;
and (d) blackbody spectra of the disk and boundary layer.}

\figcaption{The locations in the $\oms-\hs$ plane of boundary layer
solutions with $j=0.9$, 0, and -1 (solid lines) and with $\vs = -1000
\cm \pers$ (dashed line), for $\md = 10^{-4.15} \msyr$, $\ms = 0.5
\msun$, $\rs = 3 \xec$, and $\alpha = 10^{-2}$.  The locations of the
five solutions shown in Figure 3 are marked on the $j=0.9$ track;
these solutions have $\hs = 0.8$, 0.9, 1.0, 1.1, $1.2 \xec$. The
locations of the five solutions shown in Figure 4 are shown on the
$\vs = -1000 \cm \pers$ track; these solutions (from left to right)
have $j = 0.94$, 0.90, 0.84, 0.60, 0.0.}

\figcaption{Same as Fig. 1, but for five solutions along the $j=0.9$
track shown in Fig. 2, labeled by their values of $\hs$. The five
solutions have $\hs = 0.8$, 0.9, 1.0, 1.1, $1.2 \xec$, and different
values of $\vs$, and all five have $\md = 10^{-4.15} \msyr$, $\ms =
0.5 \msun$, $\rs = 3 \xec$, and $\alpha = 10^{-2}$.}

\figcaption{Same as Figs. 1 and 3, but for five solutions along the
$\vs = -1000 \cm \pers$ track in Fig. 2, labeled by their values of
$j$. The five solutions have $j = 0.94$, 0.90, 0.84, 0.60, 0.0, and
all five have $\md = 10^{-4.15} \msyr$, $\ms = 0.5 \msun$, $\rs = 3
\xec$, and $\alpha = 10^{-2}$.}

\figcaption{Same as Figs. 1, 3, and 4, but for five solutions with
$\log \alpha = -3.0$, -2.5, -2.0, -1.5, -1.0, labeled by their value
of $\alpha$. All five solutions have $\md = 10^{-4.15} \msyr$, $\ms =
0.5 \msun$, $\rs = 3 \xec$, and $j=0$.} 

\figcaption{A boundary layer and disk solution for FU Orionis. This
solution has $\md = 10^{-3.7} \msyr$, $\ms = 0.7 \msun$, $\rs = 4 \xec
\simeq 5.75 \rsun$, $j=0$, and $\alpha = 10^{-2}$. (a) shows the
angular velocity $\O$ (solid line) and the rotational velocity
$v_{rot}$ (dotted line); the Keplerian angular velocity $\ok$ is shown
for comparison (dashed line). (b) shows the effective temperature
$\te$. (c) shows cross-correlations from observed spectra (dashed
lines) and line profiles calculated from our boundary layer and disk
solution (solid lines) for $i = 60^\circ$ at 6170 \ang (left panel),
assuming $12.5 \kms$ broadening, and at 2.2 $\mic$ (right panel),
assuming $15 \kms$ broadening. (d) shows the blackbody spectrum (solid
line) and the stellar composite spectrum (dashed line) for this
solution, assuming a distance of 500 pc; the photometric data for FU
Orionis, dereddened assuming $A_V = 2.2$, are shown as square boxes.}

\figcaption{Similar to Fig. 6, but for V1057 Cygni. This solution has 
$\md = 10^{-4.0} \msyr$, $\ms = 0.5 \msun$, $\rs = 3.5 \xec \simeq
5.03 \rsun$, $j=0$, and $\alpha = 10^{-2}$. (a)-(d) are the same as in
Fig. 6, except that we have taken $i = 30^\circ$, a distance of 600
pc, and $A_V = 3.5$, and the 2.2 $\mic$ profile is broadened by $10
\kms$.}

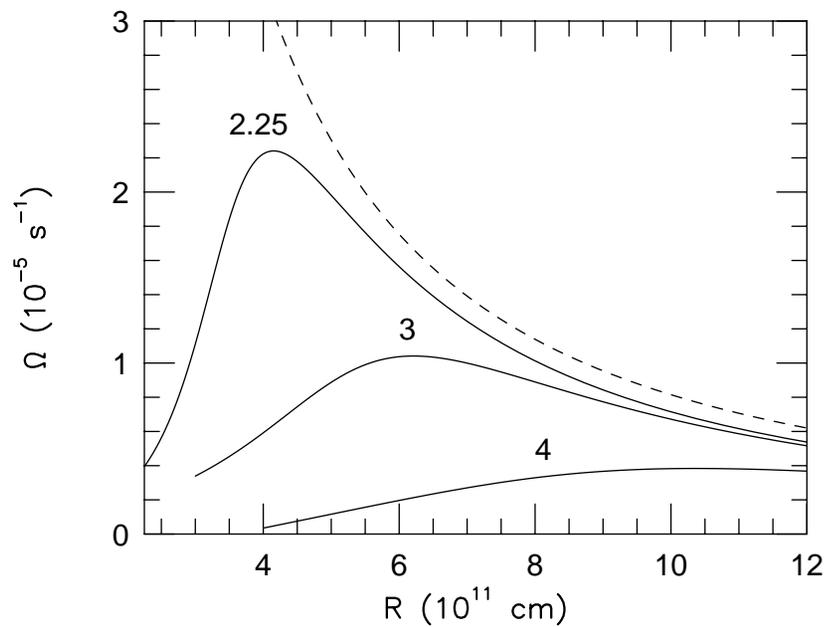
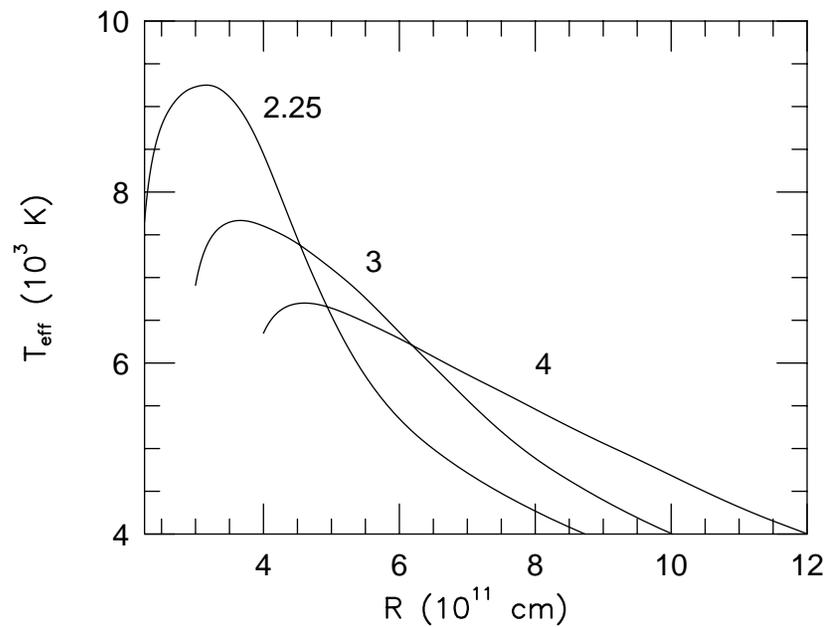
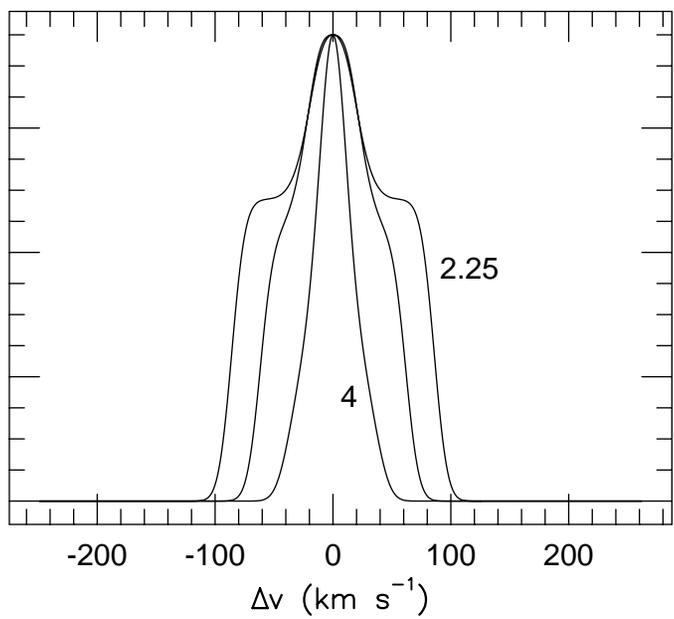
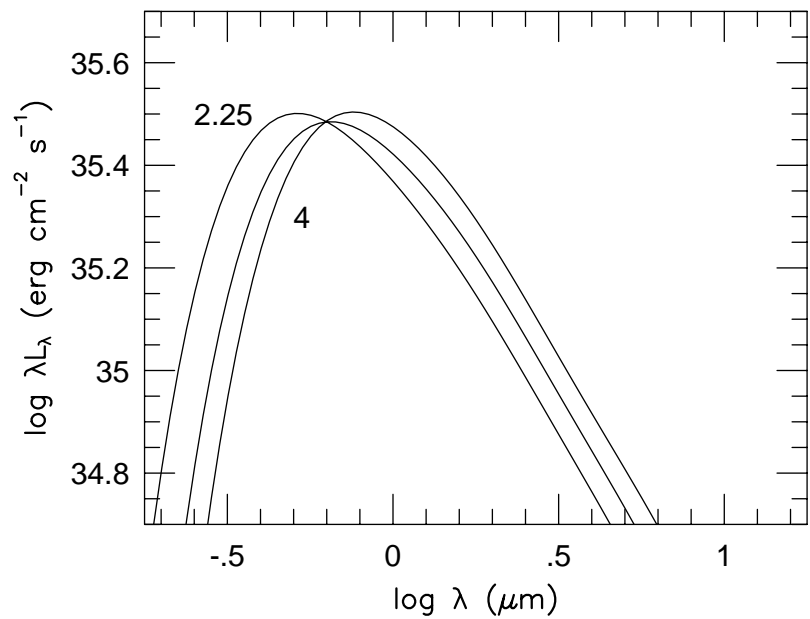
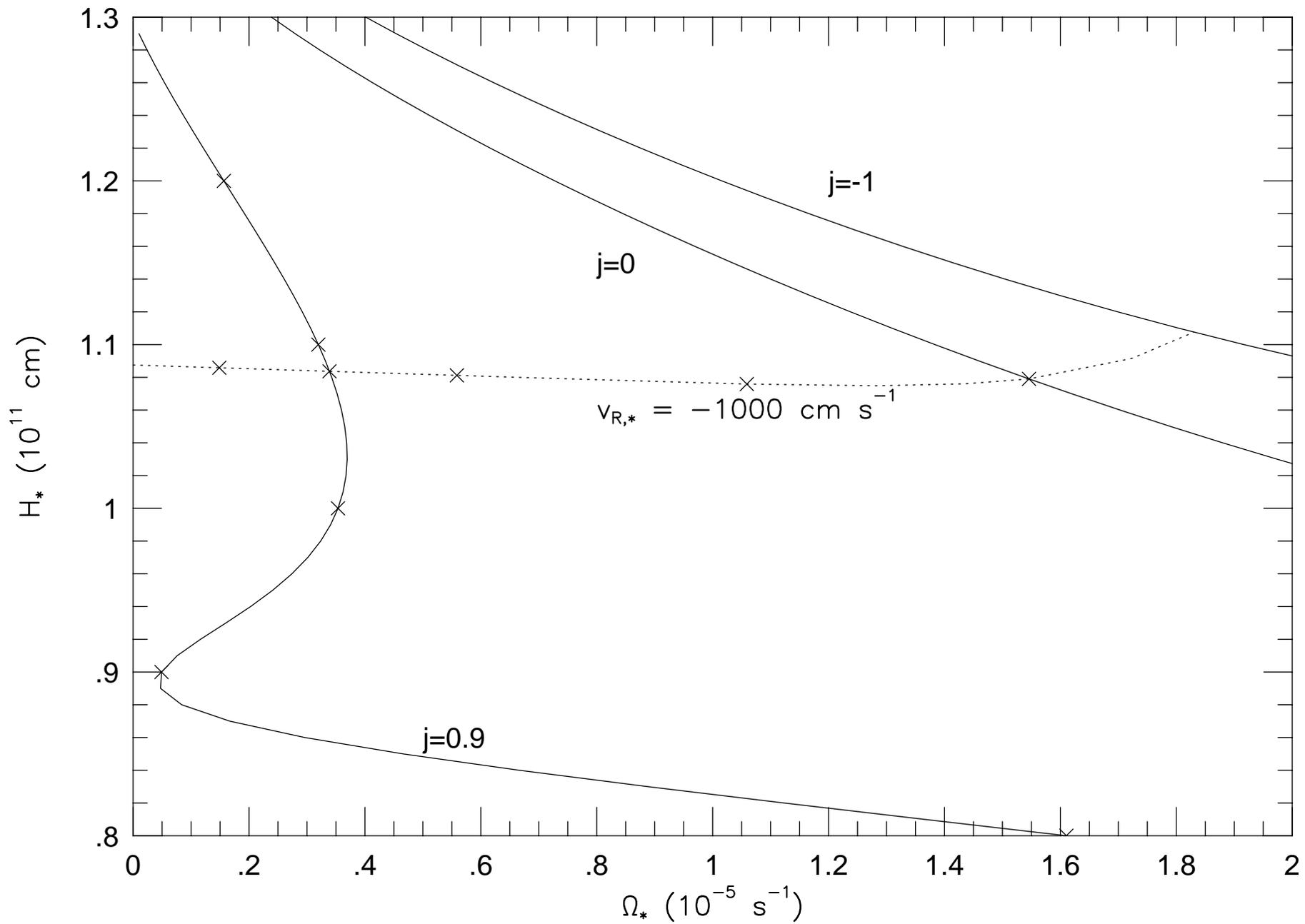
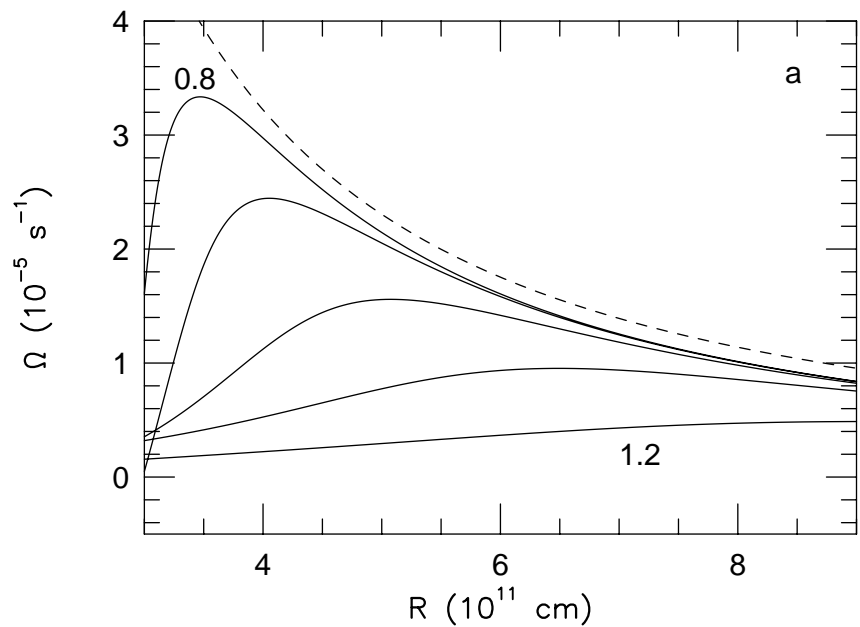
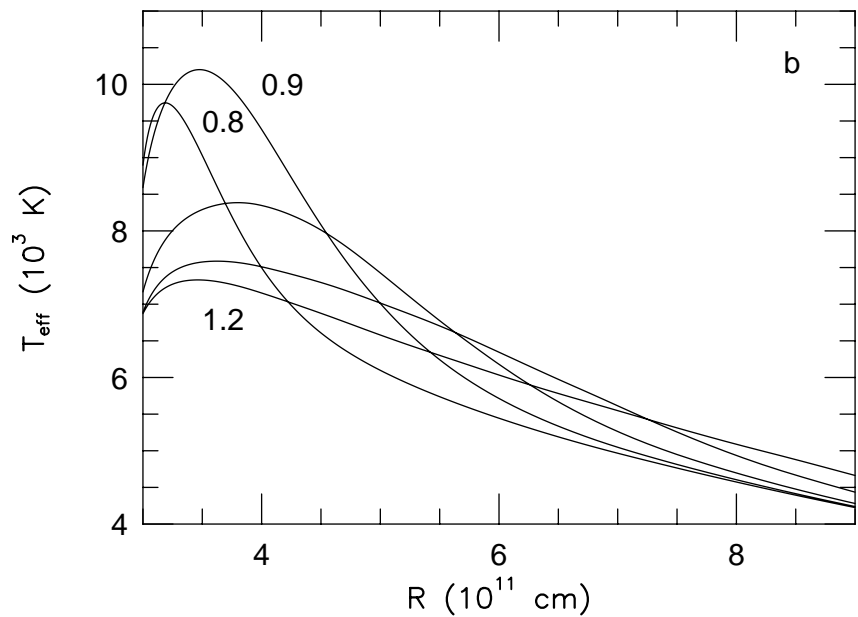
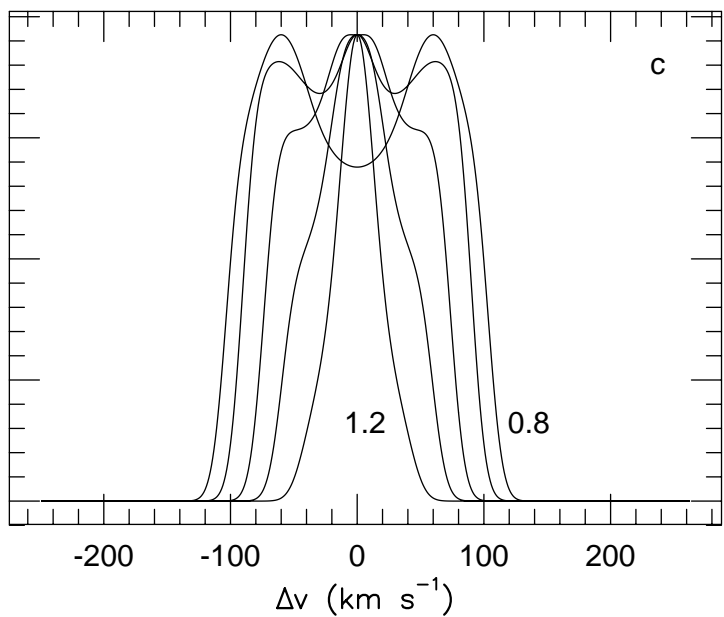
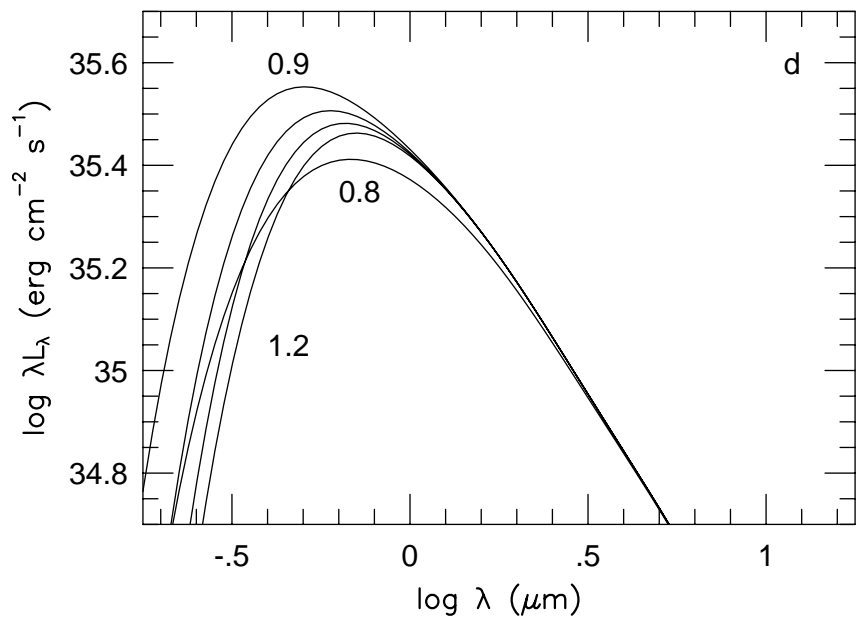
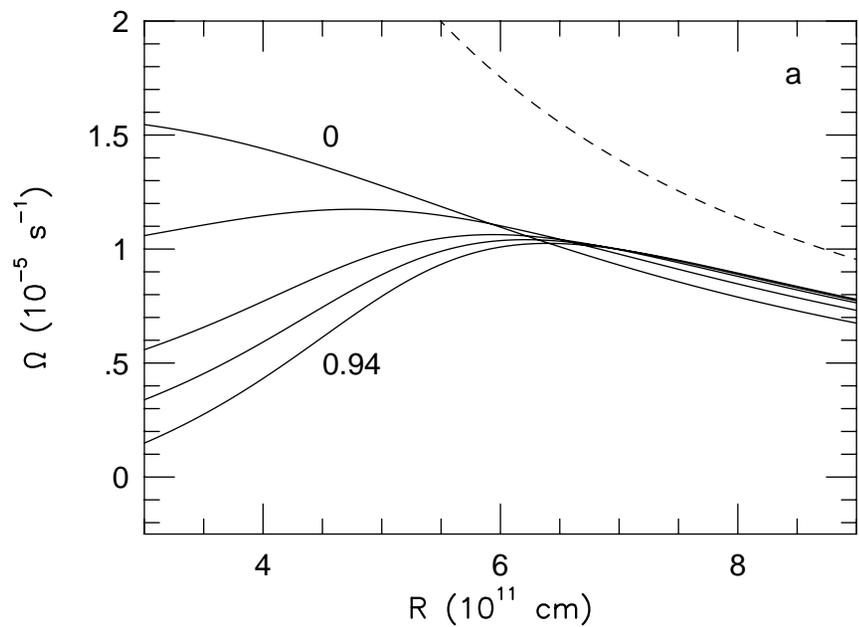
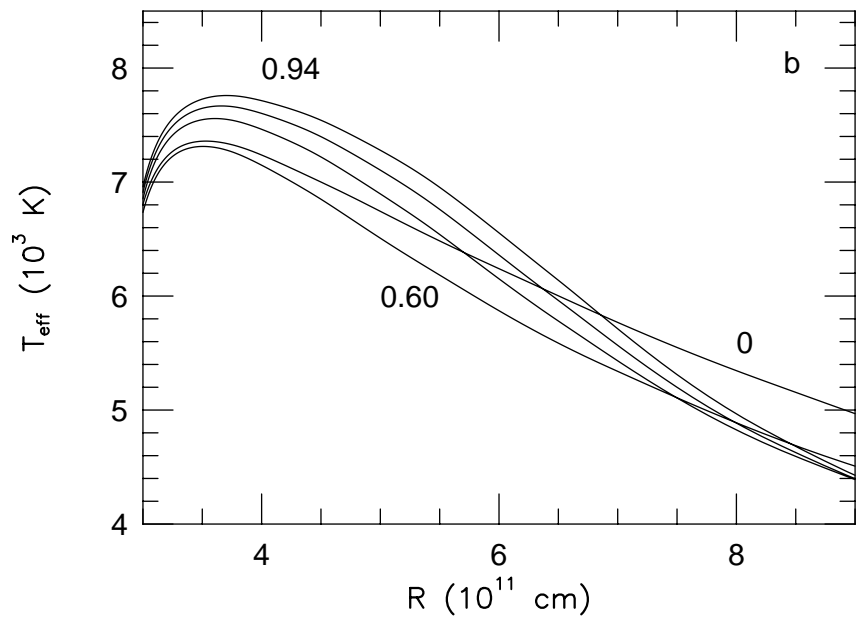
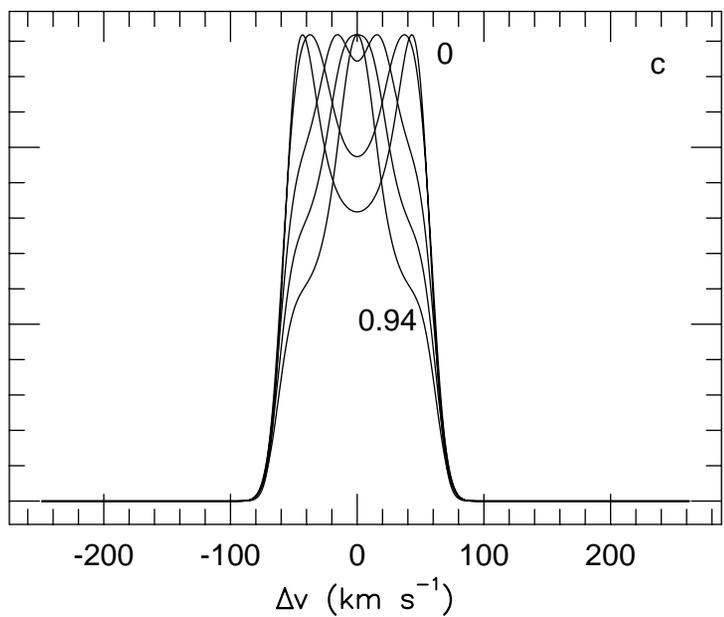
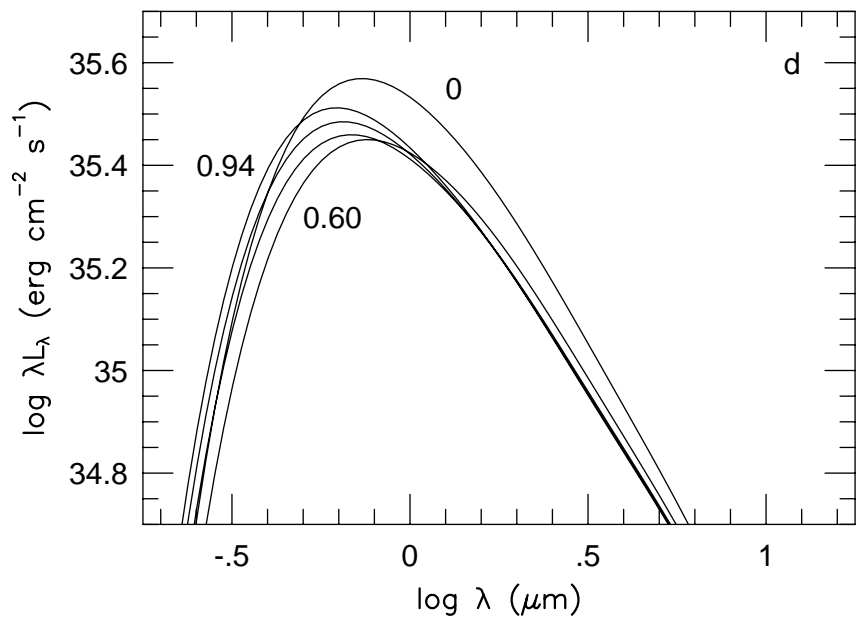
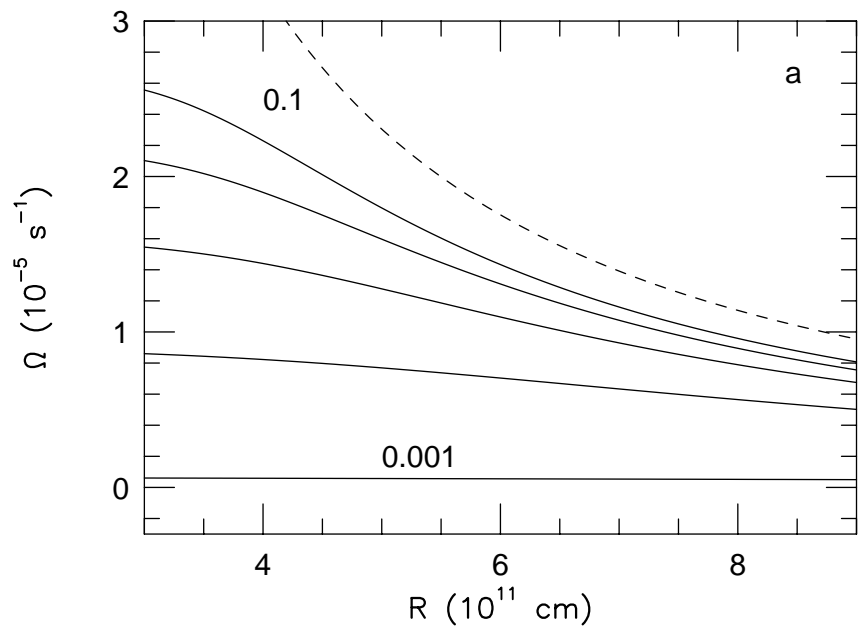
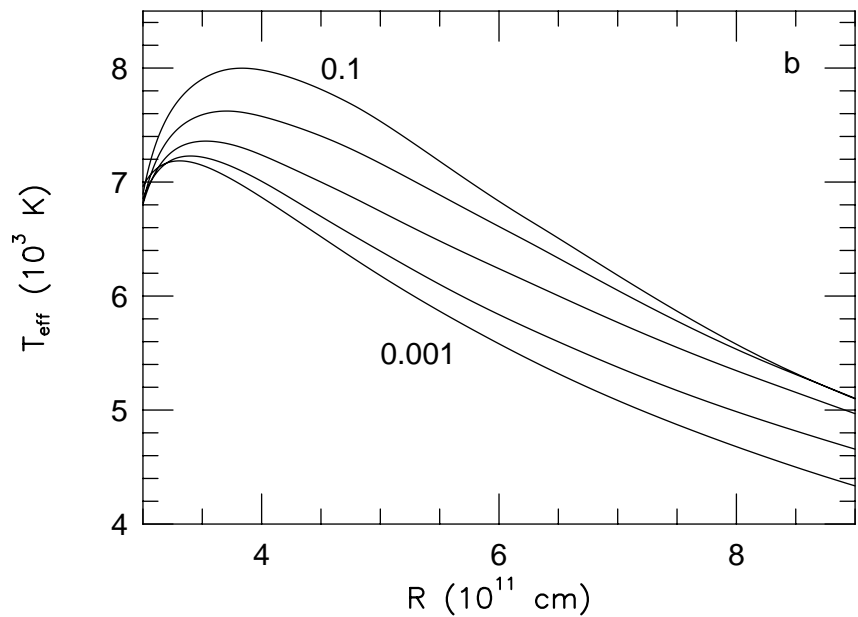
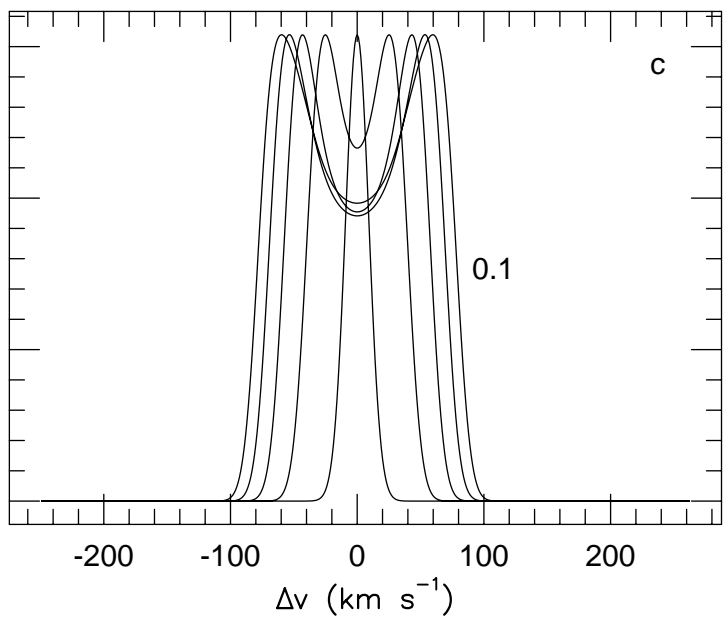
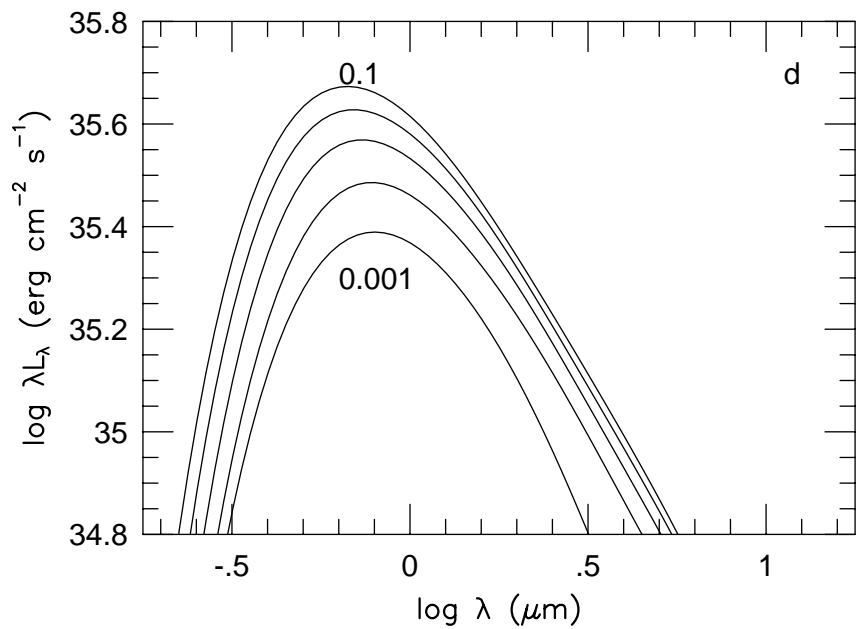
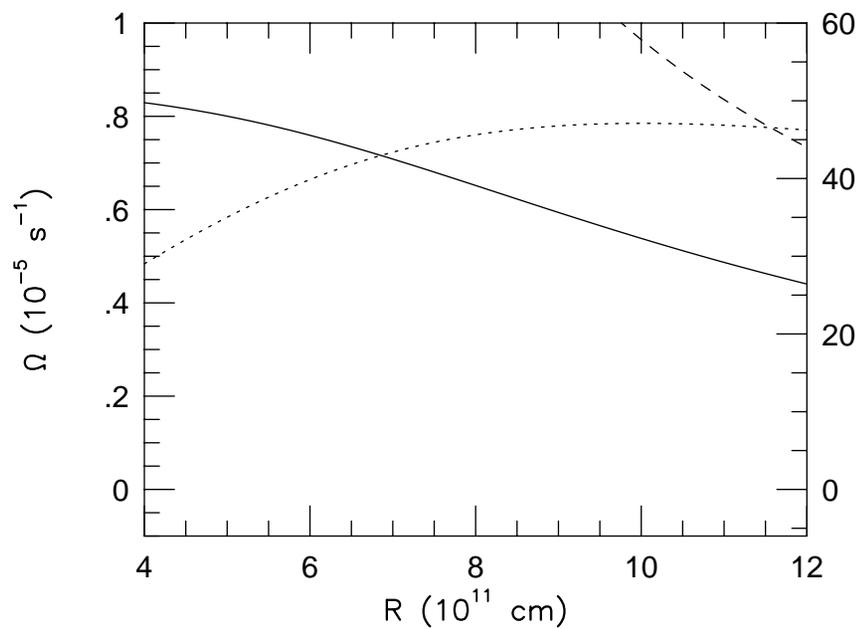
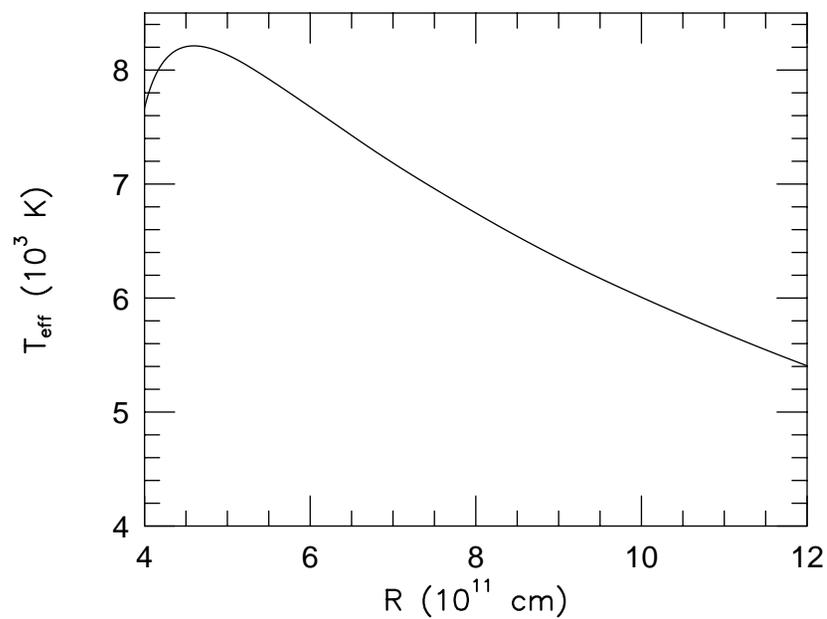
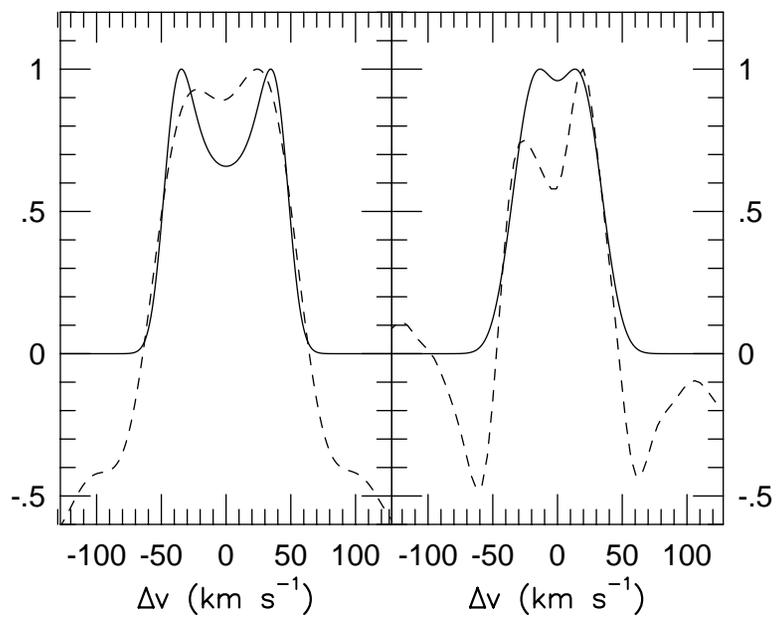
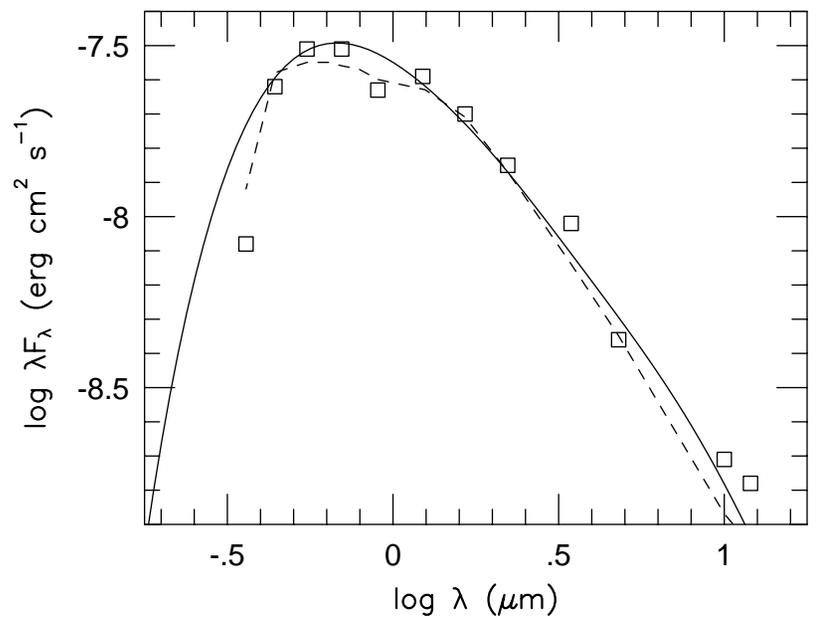
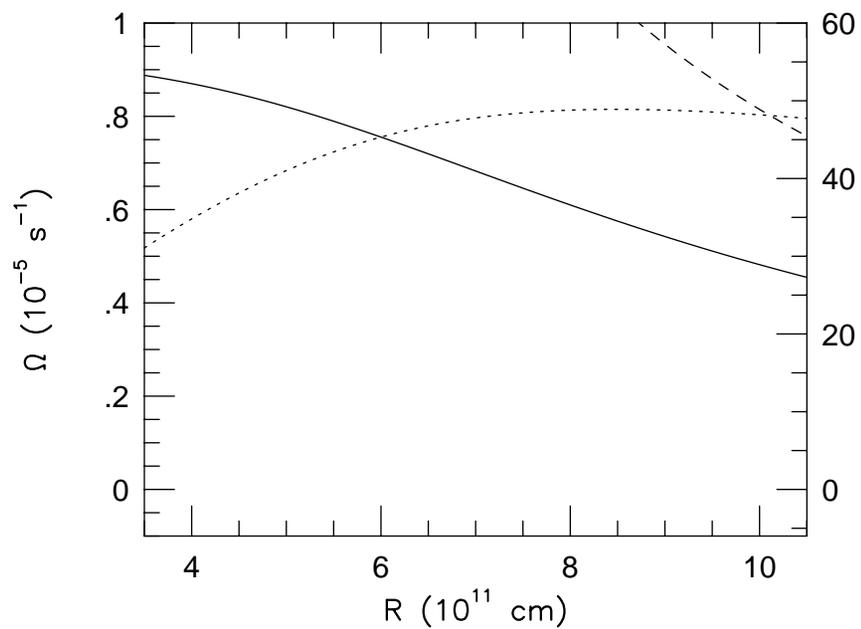
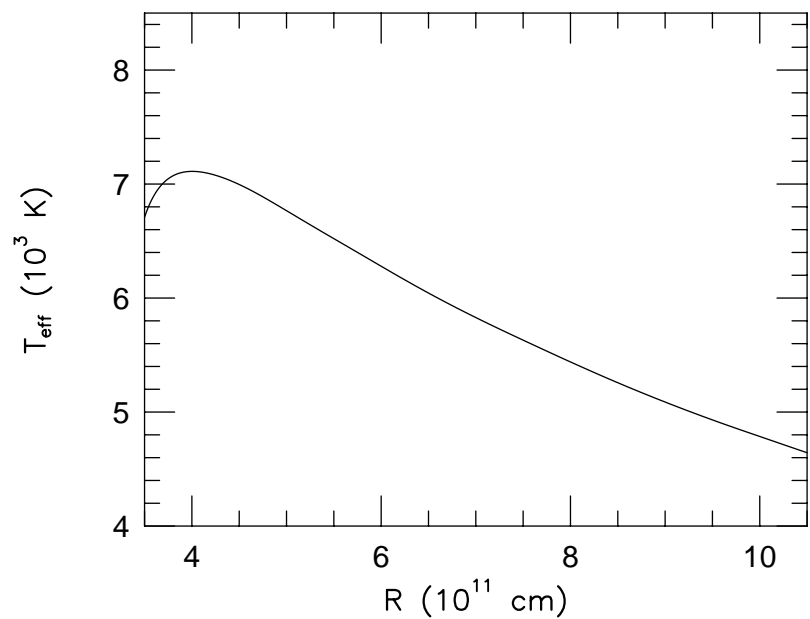
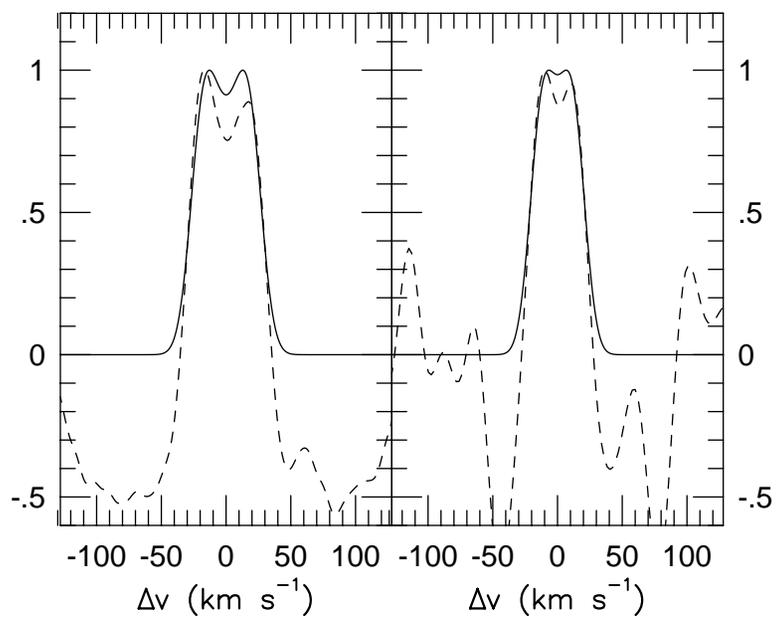
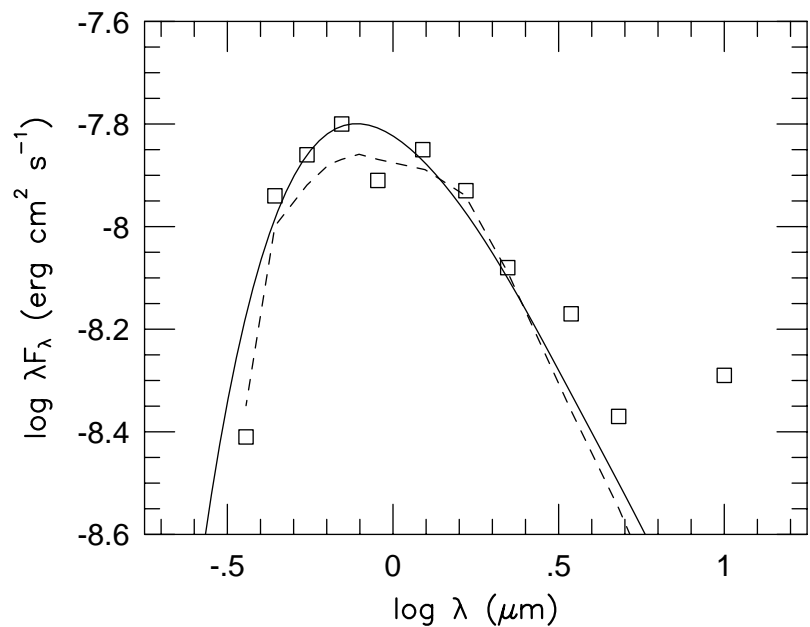
\end{document}